\documentclass[12pt]{iopart}
\expandafter\let\csname equation*\endcsname\relax
\expandafter\let\csname endequation*\endcsname\relax
\usepackage{epsfig}
\usepackage{graphicx}
\usepackage{amssymb}
\usepackage{amsmath}
\usepackage{mathtools}
\usepackage{multicol}
\usepackage [english]{babel}
\usepackage{enumitem}
\usepackage{gensymb}
\usepackage{geometry}
\usepackage{cancel}
\usepackage[titletoc,title]{appendix}
\usepackage{cite}
\usepackage{epstopdf}
\renewcommand\vec\mathbf

\begin{document}
\title{Gyrokinetic treatment of a grazing angle magnetic presheath}
\author{A Geraldini$^{1,2}$, F I Parra$^{1,2}$ and F Militello$^{2}$}
\address{$^{1}$ Rudolf Peierls Centre for Theoretical Physics, University of Oxford, Oxford, OX1 3NP, UK}
\address{$^2$ CCFE, Culham Science Centre, Abingdon, OX14 3DB, UK}
\ead{alessandro.geraldini@merton.ox.ac.uk}    
\begin{abstract}
We develop a gyrokinetic treatment for ions in the magnetic presheath, close to the plasma-wall boundary. 
We focus on magnetic presheaths with a small magnetic field to wall angle, $\alpha \ll 1$ (in radians). Characteristic lengths perpendicular to the wall in such a magnetic presheath scale with the typical ion Larmor orbit size, $\rho_{\text{i}}$. The smallest scale length associated with variations parallel to the wall is taken to be across the magnetic field, and ordered $l = \rho_{\text{i}} / \delta$, where $ \delta \ll 1$ is assumed. The scale lengths along the magnetic field line are assumed so long that variations associated with this direction are neglected. These orderings are consistent with what we expect close to the divertor target of a tokamak. We allow for a strong component of the electric field $\vec{E}$ in the direction normal to the electron repelling wall, with strong variation in the same direction. The large change of the electric field over an ion Larmor radius distorts the orbit so that it is \emph{not} circular.
We solve for the lowest order orbits by identifying coordinates, which consist of constants of integration, an adiabatic invariant and a gyrophase, associated with periodic ion motion in the system with $\alpha = \delta = 0$. By using these new coordinates as variables in the limit $\alpha \sim \delta \ll 1$, we obtain a generalized ion gyrokinetic equation. We find another quantity that is conserved to first order and use this to simplify the gyrokinetic equation, solving it in the case of a collisionless magnetic presheath. Assuming a Boltzmann response for the electrons, a form of the quasineutrality equation that exploits the change of variables is derived. The gyrokinetic and quasineutrality equations give the ion distribution function and electrostatic potential in the magnetic presheath if the entrance boundary condition is specified.
\end{abstract}

\section{Introduction}

In magnetic fusion devices, the heat flux to plasma facing components (such as divertor or limiter targets) needs to be reduced in order for these materials not to be eroded quickly. One of the well-established strategies to decrease the heat load on a divertor or limiter target is having magnetic field lines arriving at a grazing angle to it, so that the projection of the Scrape Off Layer (SOL) heat flux width $\lambda_{\text{q}}$ onto the target gets larger, increasing the area over which the heat is deposited \cite{Stangeby-book, Loarte-2007}.

\begin{figure}[h]
\centering
\includegraphics[scale=0.5]{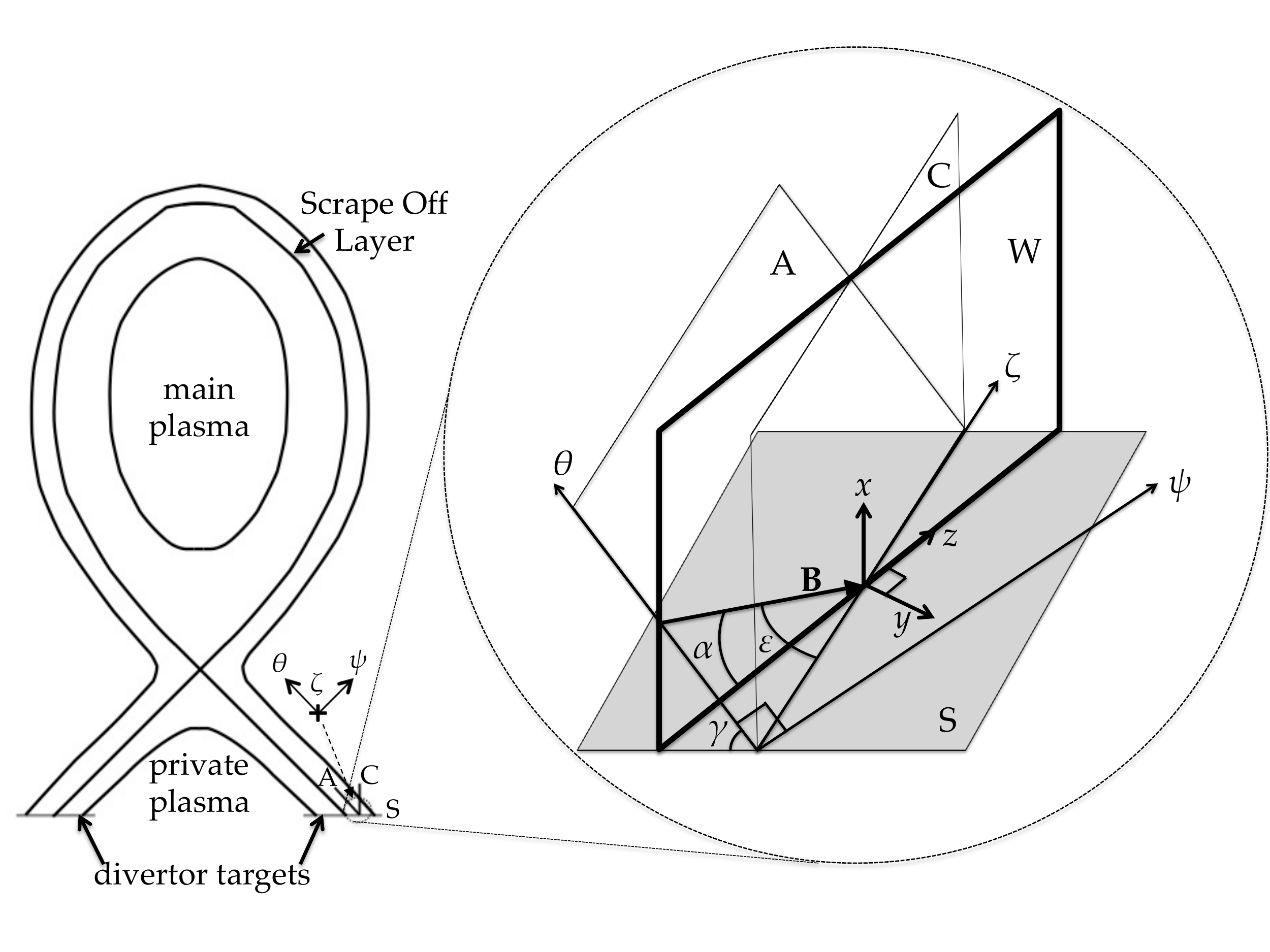}
\caption{On the left, a cartoon of the flux surface contours in the poloidal plane of a typical tokamak, with different plasma regions labelled. The region close to one of the two divertor targets (grey horizontal lines) is enlarged and shown in 3 dimensions on the right. Here, we have shown the cartesian coordinate axes $(x,y,z)$, which we use as a basis to express our equations in, and how they relate to the tokamak geometry. The divertor target (also referred to as the wall) is the grey surface S, and three planes are shown cutting through it: a plane A containing the field line $\vec{B}$ and the toroidal direction $\zeta$ (so it contains the flux surface locally), a plane C containing the $x$-axis which is locally normal to the wall and the toroidal direction, and a plane W (thicker line) containing the field line and the normal to the wall (which contains the $x$ and $z$ axes). Planes A and C are also shown on the left as black solid lines near the divertor S. In both drawings we have identified, in the region near the divertor target S, the local poloidal and toroidal axes $\theta$ and $\zeta$ respectively, and the axis locally normal to the flux surface $\psi$. On the right, we have labelled the minimum angle $\alpha$ between the field and the wall, the angle $\gamma$ between the wall and the poloidal direction, and finally the angle $\varepsilon$ between the field line and the toroidal direction.}
\label{figure-combined-divertor}
\end{figure} 

Figure \ref{figure-combined-divertor} shows the flux surface geometry in the poloidal plane of a typical tokamak (on the left) and an enlargement of the region close to a divertor target (right). %The local toroidal, poloidal and flux coordinate axes $\zeta$, $\theta$ and $\psi$ are marked in this enlarged region. 
For the rest of this paper, we will work with cartesian axes that are locally aligned with the divertor target normal (which we align with the $x$ axis) and the magnetic field $\vec{B}$ (which we restrict to the $x$-$z$ plane). Our axes are therefore aligned with the plane W as shown on the right of Figure \ref{figure-combined-divertor}. The angle that the field line makes with the wall (which has to be measured in the plane W) is labelled $\alpha$. Since tokamaks operate with magnetic field lines impinging on divertor targets at a small angle, we assume $\alpha \ll 1$ (with $\alpha$ measured in radians unless otherwise indicated) in this work.

\begin{figure}[h]
\centering
\includegraphics[width=0.5\textwidth]{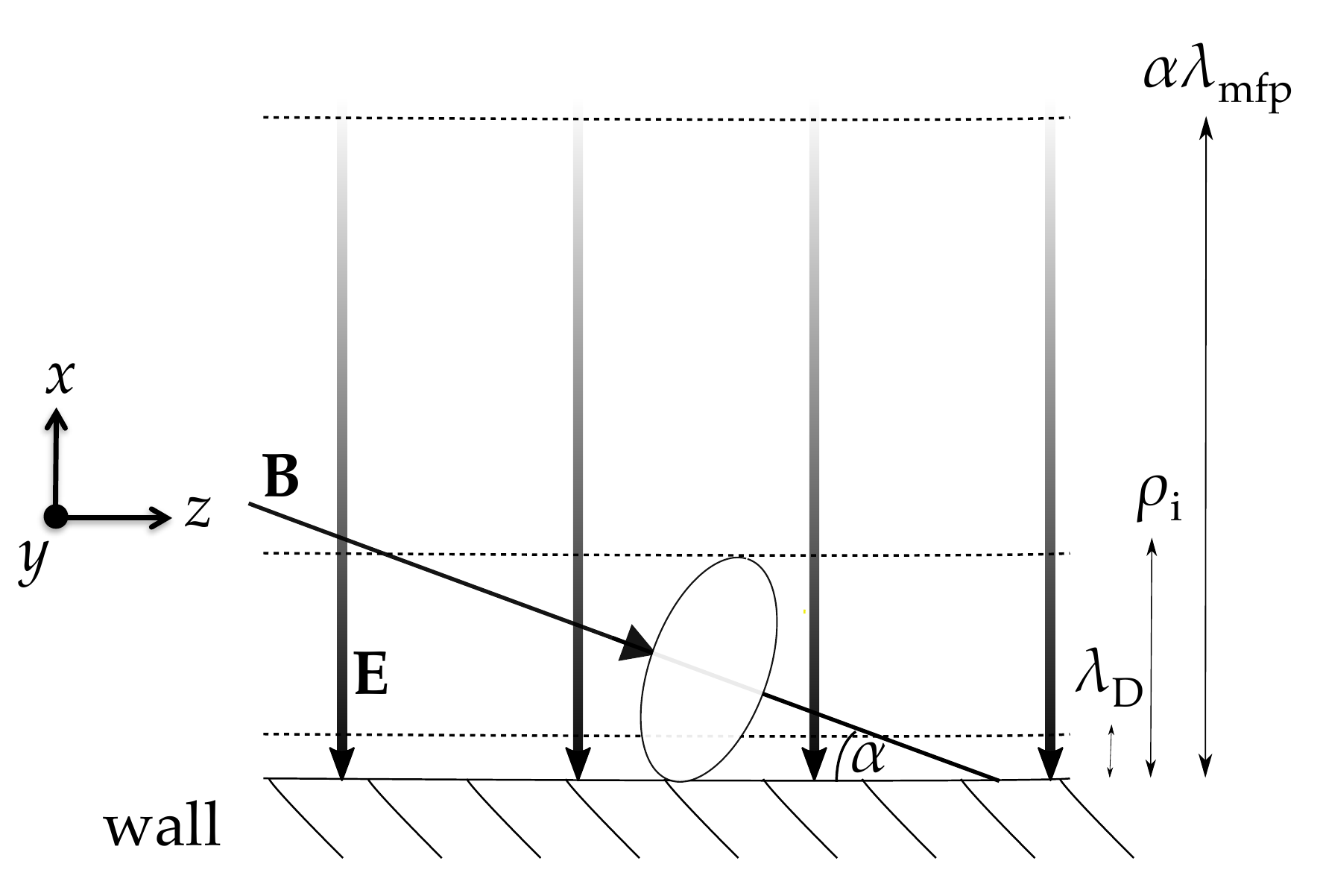}
\caption{Schematic of the different boundary layers of magnetized plasma in contact with a solid wall in the limits $\alpha \ll 1$ and $\alpha \lambda_{\text{mfp}} \gg \rho_{\text{i}} \gg \lambda_D$. The collisional layer, magnetic presheath and Debye sheath can be identified from largest to smallest. In this article, we will work with the set of axes on the left, with the $y$-axis pointing out of the page. The plane of the page in this diagram corresponds to plane W in Figure \ref{figure-combined-divertor}.}
\label{figure-boundary-layers}
\end{figure}

The plasma-wall boundary is typically split into three regions. Closest to the bulk plasma there is a collisional layer, which is several collisional mean free paths $\lambda_{\text{mfp}}$ along the magnetic field, and hence has a size of order $\alpha \lambda_{\text{mfp}}$ in the direction perpendicular to the wall. Closer to the wall a magnetic presheath, of size several ion Larmor radii $\rho_{\text{i}}$, is present \cite{Chodura-1982}. Finally the Debye sheath, several Debye lengths $\lambda_{\text{D}}$ wide, is the layer next to the wall, where quasineutrality breaks down. The separation of the layers described above works provided $\lambda_{\text{D}} \ll \rho_{\text{i}} \ll \alpha \lambda_{\text{mfp}}$, and is depicted in Figure \ref{figure-boundary-layers}. We work in the mixed limit $\lambda_{\text{D}} \ll \rho_i \sim \alpha \lambda_{\text{mfp}}$ of the quasineutral collisional magnetic presheath, neglecting collisionality only in the final parts of Sections 4 and 5. As well as modelling the plasma-wall boundary in a fusion device, it is worth mentioning that our formulation of the magnetic presheath may have applications in other areas of research, such as Hall thrusters \cite{Martinez-1998}, plasma probes \cite{Hutchinson-book} and magnetic filters \cite{Anders-1995-filters}.

In \ref{appendix:widths} we estimate the typical Debye length and ion gyroradius near the divertor target of a tokamak plasma using data from reference \cite{Militello-Fundamenski-2011} (assuming the ion temperature to be comparable to the electron temperature), and find $\lambda_{\text{D}} \sim 0.02 \text{ mm}$ and $\rho_{\text{i}} \sim 0.7 \text{ mm}$. These estimates have a weak dependence on temperature (and density in the case of the Debye length), so they are likely to be a good indication of the true relative value. The asymptotically thin Debye sheath hence appears to be a reasonable approximation. 
The collisional mean free path can also be estimated to be roughly $\lambda_{\text{mfp}} \sim 1 \text{ m}$. With a realistic value for the angle $\alpha$, $\alpha \sim 0.1$, we get $\alpha \lambda_{\text{mfp}} \sim 0.1 \text{ m}$, so our estimate indicates that the asymptotic separation between the thickness of the magnetic presheath and the thickness of the collisional layer is a good model. However, we keep the collisional term in our most general equations due to the uncertainty of our estimates associated with the width of the collisional boundary layer, and due to the important role that charge exchange collisions are believed to play (due to sputtering of the wall by neutrals) in the plasma-wall boundary. %\ref{appendix:widths} contains the data and the order of magnitude calculations used to obtain these estimates.

There is a vast literature that treats the plasma-wall boundary using fluid equations \cite{Chodura-1982, Riemann-1994, Hutchinson-2008, Stangeby-1995, Stangeby-Chankin-1995, Tskhakaya-2002, Loizu-2012, Stanojevic-2005, Zimmermann-2008}. Although, with due care, fluid equations may capture well most of the underlying physics, it is generally widely accepted that a proper kinetic treatment should be carried out to describe the boundary layer, which is kinetic in nature from the collisional layer to the wall. For example, a recent paper by Siddiqui \etal \cite{Siddiqui-Hershkowitz-2016} argues, using novel experimental measurements of ion flows in 3 dimensions near the plasma-wall boundary, that a kinetic theory of ions and neutrals is necessary in the collisional and magnetic presheaths in order to accurately predict the location and intensity of ion and charge exchanged neutral fluxes to the wall.
While some fluid approaches attempt to retain kinetic effects \cite{Chankin-1994}, there are not many fully kinetic treatments, which are usually computational in nature. Due to the difficulty in dealing with the exact kinetic equation analytically near the boundary, fewer analytic kinetic treatments exist \cite{Harrison-Thompson-1959, Riemann-review, Daube-Riemann-1999, Cohen-Ryutov-1998, Holland-Fried-Morales-1993, Daybelge-Bein-1981}. Most of the computational efforts use Particle-in-Cell (PIC) codes \cite{Tskhakaya-2004, Tskhakaya-2003, Kovacic-2009, Ahedo-2009, Khaziev-Curreli-2015}, although some use Eulerian-Vlasov advection schemes \cite{Devaux-2006, Coulette-2014, Coulette-Manfredi-2016}. In this paper we aim to develop a framework that allows more analytical insight, and that in the future will simplify the numerical computations required to solve the boundary problem in a way that captures as much of the correct physics as possible.

Our approach follows the one of Cohen and Ryutov in reference \cite{Cohen-Ryutov-1998}, who analyzed the single particle motion in the magnetic presheath geometry and introduced a set of more convenient coordinates to describe the motion. In our work, we derive an ion gyrokinetic equation for a grazing angle magnetic presheath that includes the effect of gradients parallel to the wall, and an expression for quasineutrality that exploits the new coordinates in a direct way. Moreover, we exploit an additional coordinate, which was also used in reference \cite{Holland-Fried-Morales-1993}, in order to capture the effect of the small turbulent electric field parallel to the wall in a simple way. Unlike in \cite{Holland-Fried-Morales-1993}, we do not assume that the electric field parallel to the wall is constant. 

The ion gyrokinetic equation that we derive describes the evolution of closed ion orbits in the magnetic presheath. Gyrokinetic treatments of magnetized plasmas have been developed over the last few decades in order to more effectively simulate turbulence in the core plasma \cite{Catto-1978, Lee-Myra-Catto-1983, Parra-Catto-2008}. They consist of an asymptotic expansion in the small parameter $\rho_{\text{i}} / a$, the ratio of a characteristic ion Larmor radius to the size of the device. This scale separation allows simulations over a five-dimensional phase space, instead of a six-dimensional one, by effectively averaging over gyro-orbit timescales that are much shorter than all other timescales. The problem with applying conventional gyrokinetics in the magnetic presheath is that typical gradient scale lengths of the lowest order quantities  in the direction perpendicular to the wall are comparable to the characteristic ion Larmor radius. Therefore, to treat the ions kinetically at the magnetic presheath scale, we develop a modification to the gyrokinetic formalism that makes it valid in the magnetic presheath by retaining orbit distortion to lowest order. Our model of the magnetic presheath assumes a grazing angle magnetic field, $\alpha \ll 1 $, and includes small gradients parallel to the wall, $\delta = \rho_{\text{i}} / l  \ll 1$, where $l$ is the characteristic length scale parallel to the wall. The ordering for the small electric fields parallel to the wall is consistent with fields expected due to turbulent structures impinging on the divertor target. The variables that we use are the constants of the ion motion in the system with $\alpha = 0$ (magnetic field parallel to wall) and $\delta = \rho_{\text{i}} / l = 0$ (no gradients parallel to the wall), and a gyrophase that is generalized to account for orbit distortion.  We deduce the new gyrokinetic equation by asymptotically expanding the ion motion in the two small parameters $\alpha$ and $\delta$. 

The paper is structured as follows. After explaining how the different quantities are ordered in Section 2, we analyse the ion single particle motion in the magnetic presheath in Section 3. There, we set $\alpha = 0$ and $\delta = 0$ exactly, focusing on the problem of a strongly varying electric field exactly perpendicular to a constant magnetic field, which we call the zeroth order problem. This problem is analysed only in the context of ion trajectories, \emph{not} in the context of plasma-wall interaction with a magnetic field parallel to the wall. We derive some orbit parameters (which are constants of integration of the zeroth order problem) and, after identifying the conditions for periodicity of the ion motion, an expression for the gyrophase of the orbit. We also introduce an adiabatic invariant that was derived in reference \cite{Cohen-Ryutov-1998}.

The orbit parameters become slowly changing variables in Section 4, where we consider a layer in which the electric and magnetic fields are almost but not exactly perpendicular to one another, as is the case in a grazing angle magnetic presheath. We begin Section 4 by writing the ion kinetic equation, then change variables by exploiting the slowly changing orbit parameters and gyrophase which we introduced in Section 3. Such a change of variables allows us to show that the lowest order distribution function is gyrophase independent, and to write a kinetic equation averaged over gyrophase (a gyroaveraged kinetic equation, or gyrokinetic equation) to next order, which encodes variation over the typical magnetic presheath timescale. We simplify this gyrokinetic equation further by introducing another orbit parameter, which is a constant of the motion to first order. Changing to a set of variables that includes the new orbit parameter and the adiabatic invariant, the collisionless magnetic presheath ($\rho_{\text{i}} \ll  \alpha \lambda_{\text{mfp}} $) has a simple solution for the distribution function in terms of the boundary condition at its entrance.

In Section 5 we derive a quasineutrality condition that employs the new variables, assuming Boltzmann distributed electrons. The quasineutrality equation is then simplified by taking the limit of a collisionless magnetic presheath $\rho_{\text{i}} \ll \alpha \lambda_{\text{mfp}}$. For such a simplified presheath, the quasineutrality equation may be solved iteratively to find the self-consistent electrostatic potential that develops in the magnetic presheath. In the final part of Section 5, we make some imporant remarks regarding the validity of the equations presented in this paper.

We conclude in Section 6 by summarizing our results, discussing the strenghts and limitations of our gyrokinetic treatment, and outlining plans for future work.

\section{Orderings}
\begin{figure}[h]
\centering
\includegraphics[scale = 0.4]{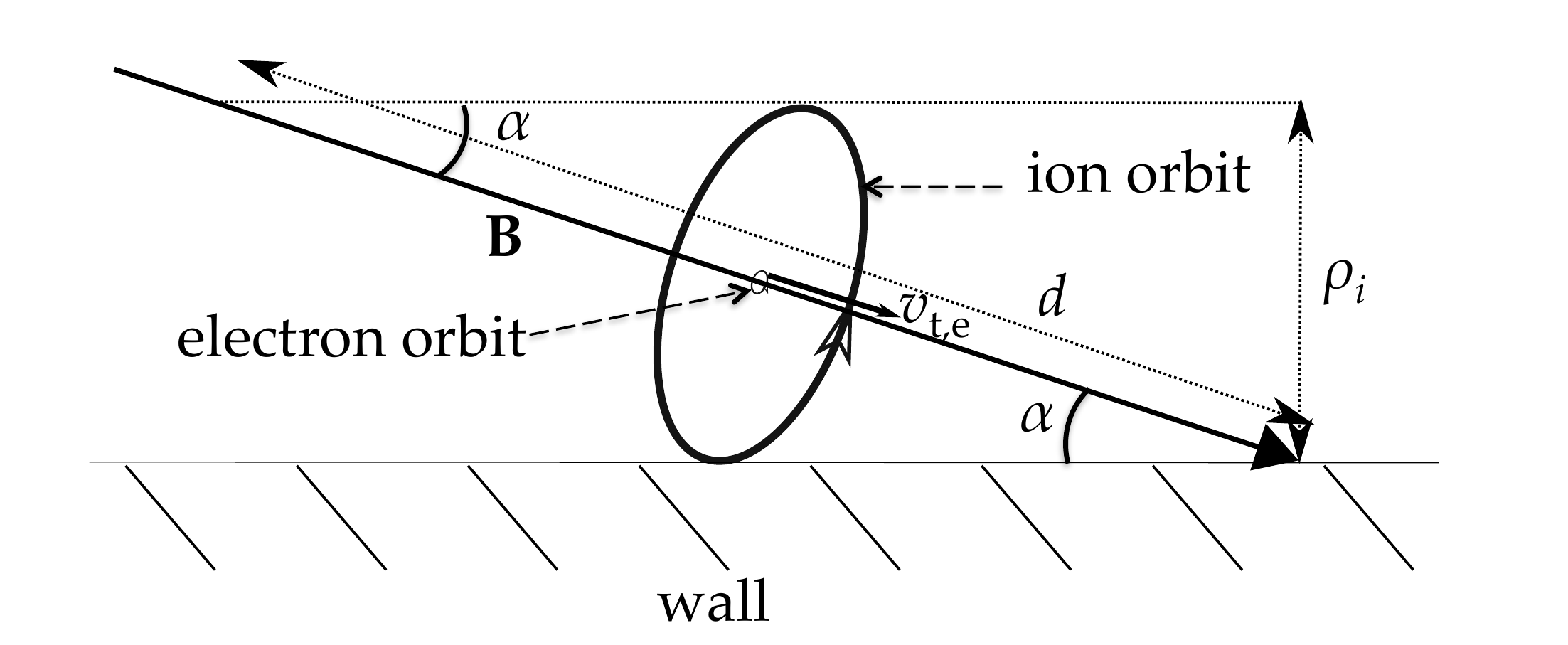}
\caption{Electrons and ions travelling towards the wall orbiting around a grazing angle open field line. It takes a similar time for ions and electrons to reach the wall if 
$d / v_{\text{t,e}} \sim \rho_{\text{i}} / v_{\text{t,i}}$, where $v_{\text{t,e}} = \sqrt{2T_{\text{e}}/m_{\text{e}}}$ and $v_{\text{t,i}} = \sqrt{2T_{\text{i}}/m_{\text{i}}}$ are the electron and ion thermal speeds, $T_{\text{e}}$ and $T_{\text{i}}$ are the electron and ion temperatures, $m_{\text{e}}$ and $m_{\text{i}}$ are the electron and ion masses. Then,
$\alpha \simeq \rho_{\text{i}} / d \sim v_{\text{t,i}} / v_{\text{t,e}}$,
 which simplifies to $\alpha \sim \sqrt{m_{\text{e}} / m_{\text{i}}}$
  if $T_{\text{e}} \sim T_{\text{i}}$. For the wall to be negatively charged, we need the electrons to reach it faster than the ions, so we require $d / v_{\text{t,e}}\ll \rho_{\text{i}} / v_{\text{t,i}}$ which leads to $\alpha \gg \sqrt{m_{\text{e}} / m_{\text{i}}}$. }
\label{figure-limiting-angle}
\end{figure}

In this paper, we denote the electric and magnetic fields in the magnetic presheath as $\vec{E}$ and $\vec{B}$ respectively, and use the coordinate axes in Figure \ref{figure-boundary-layers}. The ordering for the angle with which the magnetic field impinges on the wall is $\alpha \ll 1$. Throughout this work we assume a negatively charged, electron repelling wall, valid provided that the time it takes for the electrons to reach the wall is shorter than the time it takes for the ions. Close to the wall in the magnetic presheath, the time it takes for an ion to intersect the wall is a typical ion gyroperiod, while the time it takes for an electron (due to its much smaller Larmor radius) is given by its faster streaming along the field line towards the wall, as shown in Figure \ref{figure-limiting-angle}. The criterion that must be satisfied for electrons to reach the wall faster, leading to a negatively charged wall, is $\alpha \gg \sqrt{m_{\text{e}} /m_{\text{i}}} \simeq 0.02 \left( \simeq 1^{\circ} \right) $ \cite{Coulette-Manfredi-2016, Stangeby-2012}, where $m_{\text{e}}$ and $m_{\text{i}}$ are the electron and ion mass respectively, and the estimate is made using the mass of a deuterium ion. Hence, we assume 
\begin{align}  \label{ordering-angle-range}
\sqrt{\frac{m_{\text{e}}}{m_{\text{i}}}} \ll \alpha \ll 1 \text{.}
\end{align}
In current tokamaks, the angle $\alpha$ usually lies in the range $4^{\circ}-11^{\circ}$, therefore $\alpha \sim 0.07-0.2$ and the ordering (\ref{ordering-angle-range}) is approximately satisfied. In ITER, it is expected that $\alpha$ will be around $2.5^{\circ}$ \cite{Pitts-2009}, which means that divertor targets may become ion repelling, in which case our model of the magnetic presheath would break down. 

The characteristic lengths perpendicular to the wall are set by the thickness of the magnetic presheath, which is of the order of the ion gyroradius $\rho_{\text{i}}$. The characteristic lengths parallel to the wall are most likely constrained by the size of the turbulent structures in the SOL, which are assumed much larger than $\rho_{\text{i}}$.
The $z$ direction is mostly along the magnetic field, a direction in which turbulent structures are elongated, while the $y$ direction is mostly across the magnetic field. From this, we argue in \ref{appendix:turbulence} that we expect gradients in the $z$ direction to be ordered smaller than in the $y$ direction.
We take the orderings 
\begin{align} \label{xyz-order}
x \sim \rho_{\text{i}} \ll y \sim l \ll z \sim \min\left(l/\alpha , l / \delta \right) \text{}
\end{align}
for the length scales associated with the different coordinate directions,
where $l$ is the characteristic cross-field size of turbulent structures in the SOL and $\delta = \rho_{\text{i}} / l \ll 1$ is a small parameter relating the different length scales. From reference \cite{Carralero-2015}, we estimate $l \sim 10 \text{ mm}$ in a typical tokamak, which leads to $\delta \sim 0.07$, which is indeed small. To be flexible, we take the maximal ordering 
\begin{align} \label{max-ord}
\alpha \sim \delta
\end{align}
throughout most of this work. At the end of Section 5, we write the quasineutrality equation for the special case of a simpler system where gradients parallel to the wall are completely neglected, corresponding to the subsidiary expansion $\delta \ll \alpha$. 

The individual ion velocity is ordered
\begin{align}
\left| \vec{v} \right| \sim v_x \sim v_y \sim v_z  \sim v_{\text{t,i}} = \sqrt{\frac{2T_{\text{i}}}{m_{\text{i}}}} \text{,}
\end{align}
where $v_{\text{t,i}}$ is the ion thermal speed, $T_{\text{i}}$ is the ion temperature and $m_{\text{i}}$ is the ion mass. Note that $v_x$ and $v_y$ are predominantly associated with the orbital motion (and the $\vec{E} \times \vec{B}$ drift, as we will see) of the particle, while $v_z$ is the parallel streaming velocity to a very good approximation. We order the ion and electron temperatures to be of similar size, $T_{\text{i}} \sim T_{\text{e}} \sim T$.

We make two assumptions which we justify in the last few paragraphs of this section. Firstly, we assume that the magnetic fields produced by currents in the magnetic presheath plasma are so small compared to the external magnetic field that they can be neglected. Secondly, we assume that the plasma in the magnetic presheath is electrostatic, which allows us to define the electrostatic potential $\phi$ via $\vec{E} = - \nabla \phi$. We will also treat the external magnetic field $\vec{B}$ as constant in space and time. This is justified because the length scale of $\vec{B}$ is set by the curvature of the device, which is typically much larger than $l$, while the time variations of this field are also expected to be negligible.

The electrostatic potential changes in the magnetic presheath are allowed to be
   \begin{align} \label{order-pot}
\phi \sim \frac{T}{e} \text{.}
   \end{align} 
The ordering (\ref{order-pot}) is consistent with the potential drop across the magnetic presheath being $\sim (T/e) \ln \alpha$, a scaling that was first predicted by Chodura using fluid equations \cite{Chodura-1982}. The size of the electric field components follows from the ordering of the potential (\ref{order-pot}) and the length scales (\ref{xyz-order}), 
\begin{align} \label{order-E}
\frac{\partial \phi}{\partial z}  \sim \frac{\delta T}{el} \sim \frac{\alpha T}{el} \ll  \frac{\partial \phi}{\partial y} \sim \frac{T}{el} \ll \frac{\partial \phi}{\partial x} \sim \frac{T}{e \rho_{\text{i}}} \text{.}
\end{align}

The typical gyrofrequency of an ion orbit in the magnetic presheath is $\Omega = ZeB / m_{\text{i}} \sim v_{\text{t,i}} / \rho_{\text{i}}$, with $Z \sim 1$ the ion atomic number and $e$ the proton charge. The strong electric field normal to the wall leads to a thermal $\vec{E} \times \vec{B}$ drift in the $y$ direction, $(1 / B) \partial \phi / \partial x \sim v_{\text{t,i}}$. %Therefore, flow along $y$ is ordered $v_{\text{t,i}}$. 
Because potential gradients in this direction are small, the drifting particles will be exposed to significant potential changes over a timescale much longer than the orbital one. This means that the effect of this drift is unimportant to lowest order.
The electric field component $\partial \phi / \partial y$ parallel to the wall leads to an $\vec{E} \times \vec{B}$ drift in the $x$ direction, normal to the wall, that is first order in $\delta$. Therefore, ions drift towards or away from the wall at a speed $\sim \delta v_{\text{t,i}}$. This drift competes with the projection of the parallel flow towards the wall $\sim \alpha v_{\text{t,i}}$ when we take the maximal ordering $\delta \sim \alpha$, consistent with previous work \cite{Loizu-2012}. 
Parallel streaming and the presence of an absorbing wall leads to an expected ion flow $\sim v_{\text{t,i}}$ in the $z$ direction. This is also expected from the fluid Chodura condition, which states that ion flow parallel to the magnetic field must be at least sonic when entering the magnetic presheath \cite{Chodura-1982}. From (\ref{order-E}), potential changes due to motion in the $z$ direction happen over a timescale so much longer than the orbital timescale that we may neglect them even to first order.

The magnetic presheath has a size $\rho_{\text{i}}$. Considering that the drift in the $x$ direction is of order $\delta v_{\text{t,i}} \sim \alpha v_{\text{t,i}}$, we expect the characteristic time $t_{\text{MPS}}$ that it takes for an ion to reach the wall after having entered the magnetic presheath to be $\rho_{\text{i}} / \delta v_{\text{t,i}} \sim \rho_{\text{i}} / \alpha v_{\text{t,i}} $,
\footnote{In Figure \ref{figure-limiting-angle}, we considered the time that it takes for an ion to reach the wall once its gyro-orbit touches the wall. Here, we are considering the time that it takes for a gyro-orbit to drift to the wall in the first place. }
which becomes 
\begin{align} \label{tMPS}
t_{\text{MPS}} \sim \frac{1}{\Omega \delta} \sim \frac{1}{\Omega \alpha} \text{.}
\end{align}
The gyrokinetic equation derived in Section 4 is accurate to variations occurring over this timescale. The size of the time derivative $\partial / \partial t$ is set by the turbulence in the SOL, and is given by (see \ref{appendix:turbulence}) 
\begin{align} \label{tcorr}
 \frac{\partial}{\partial t} \sim  \delta^2 \Omega \text{.}
 \end{align}
Because this partial derivative is higher order compared to $1/t_{\text{MPS}}$, it does not appear in the first order gyrokinetic equation of Section 4. From our earlier discussion about potential changes seen by a particle due to its parallel streaming, the same can be said about the term $v_z \partial / \partial z$ that encodes changes due to the very small gradients in the $z$ direction.
    
    Ions and electrons $\vec{E} \times \vec{B}$ drift in the same direction, so their contributions to the current partially cancel each other. However, because ions have a large Larmor orbit, they experience a strongly varying field over an orbit. Therefore, their $\vec{E}\times \vec{B}$ drift in the $y$ direction can differ from the electron one substantially, which leads to a large current density $j^D_y \sim en_{\text{i}}v_{\text{t,i}}$, where $n_{\text{i}}$ is the ion density, in this direction. The ``D'' superscript denotes current that is produced by the particle drifts in the plasma.
We can arrive at this estimate for $j_y^D$ by analyzing the size of diamagnetic ion and electron flows parallel to the wall \cite{Chankin-1994}. The order of magnitude of the diamagnetic current in the $y$ direction is $\left(1/ B^2\right) \left( \vec{B} \times \nabla p \right)_y \sim \left(1/B\right)\partial p / \partial x \sim en_{\text{i}} v_{\text{t,i}}$, where $p \sim m_{\text{i}} n_{\text{i}} v_{\text{t,i}}^2$ is the plasma pressure. From the ordering of the plasma flow in the $x$ direction, the size of the current normal to the wall is expected to be $j_x^D \sim \delta en_{\text{i}}v_{\text{t,i}} \sim \alpha e n_{\text{i}} v_{\text{t,i}}$. Again, we can obtain this using the component of the diamagnetic current in the $x$ direction, $\left(1/ B^2\right) \left( \vec{B} \times \nabla p \right)_x \sim \left(1/B\right)\partial p / \partial y \sim \delta en_{\text{i}} v_{\text{t,i}}$.

We proceed to demonstrate that the neglect of magnetic fields produced by magnetic presheath currents and the electrostatic assumption are both justified.
For the remainder of this section (and in \ref{appendix:largejz}), we refer to the constant externally produced field as $\vec{B}^{c}$. We have, from our choice of axes (see Figure \ref{figure-boundary-layers}), $B_x^{c} = - B^{c} \sin \alpha \sim \alpha B^c$, $B_y^{c} = 0$, $B_z^{c} = B^c \cos \alpha \sim B^c$. The plasma current $\vec{j}^D$ in the boundary layer can produce a magnetic field $\vec{B}^p$.
Using (\ref{xyz-order}), $\nabla \cdot \vec{B^{p}} = 0$ gives
\begin{align} \label{ordering-Bplasma}
B^{p}_x  \sim \delta^2 B^{p} \sim \delta \alpha B^{p} \ll B_y^{p} \sim \delta B^{p} \sim \alpha B^{p} \ll B_z^{p} \sim B^{p} \text{.}
\end{align}
Amp\`ere's law is
\begin{align} \label{Ampere}
\mu_0 \vec{j}^D = \nabla \times \vec{B^{p}} \text{,}
\end{align}
where $\mu_0$ is the vacuum permeability. 
Using (\ref{xyz-order}) and (\ref{ordering-Bplasma}) to order the RHS of (\ref{Ampere}), we obtain the orderings
\begin{align} \label{j-orderings}
j_x^D \sim j_z^D \sim \frac{B^p}{\mu_0 l} \ll j_y^D \sim \frac{B^p}{\mu_0 \rho_{\text{i}}} \text{.}
\end{align}
The earlier orderings for the current deduced from particle motion ($j_x^D \sim \delta en_{\text{i}} v_{\text{t,i}} \sim \alpha en_{\text{i}} v_{\text{t,i}}$ and $j_y^D \sim en_{\text{i}} v_{\text{t,i}}$) are consistent with equation (\ref{j-orderings}) if we take $j_z^D \sim j_x^D \sim \delta en_{\text{i}} v_{\text{t,i}} \sim \alpha en_{\text{i}} v_{\text{t,i}}$. This ordering is consistent with what we would expect from the piece of the parallel current that is produced in response to the perpendicular currents produced by particle drifts (an analogue of the Pfirsch-Schluter current \cite{Hirshman-1978-Pfirsch-Schluter}).\footnote{This does not imply, as discussed in the last paragraph of Section 2 and in \ref{appendix:largejz}, that larger parallel currents cannot be present in the magnetic presheath.}
From these estimates of the currents, it follows that 
\begin{align} \label{ordering-beta}
\frac{B^{p}}{B^c} \sim \beta \ll 1 \text{,}
\end{align}
where $\beta = 2 \mu_0 p / (B^{c})^2$ is the plasma beta parameter. This parameter is typically small in the core and is even smaller in the SOL ($\beta \sim 0.004$ inferred from reference \cite{Militello-Fundamenski-2011}), so that the field produced by the plasma in the magnetic presheath is much smaller than the externally generated one. 

In order to neglect the plasma produced magnetic field in our equations, we require each component of it to be negligible compared to either the respective component or the smallest retained component of the external magnetic field $\vec{B}^c$. Considering the non-zero components of $\vec{B}^c$ (the $z$ and $x$ components), we require $B^p_z \sim B^p \ll B_z^c \sim B^c$ and $B_x^p \sim \delta \alpha B^p \ll B_x^c \sim \alpha B^c$, which are both satisfied if the inequality (\ref{ordering-beta}) holds. In addition to this we require that $B_y^p \ll B_x^c$ (because $B^c_x$ is the smallest retained component of the external magnetic field), which is satisfied if (\ref{ordering-beta}) holds. This discussion justifies taking $\vec{B} = \vec{B}^c = \text{constant}$ in our equations and hence neglecting all plasma produced magnetic field components.

The electrostatic approximation is valid if each component of the \emph{non-electrostatic} piece, $\vec{E}^p$, of the electric field (which is induced by the plasma produced magnetic fields) is negligible compared to either the respective component or the smallest retained component of the \emph{electrostatic} piece, $-\nabla \phi$, of the electric field. The smallest retained component of the electric field is $\partial \phi / \partial y \sim T/el$ because we will neglect $\partial \phi / \partial z$ as discussed earlier. With this consideration and using (\ref{order-E}), we require $E_x^p \ll T / e\rho_{\text{i}} \sim v_{\text{t,i}} B^c $, $E_y^p \ll T / el \sim \delta v_{\text{t,i}} B^c $ and $E_z^p \ll T / el \sim \delta v_{\text{t,i}} B^c $ in order to justify the electrostatic approximation. The induction equation is
\begin{align} \label{induction}
\frac{\partial \vec{B}^{p}}{\partial t} = - \nabla \times \vec{E}^{p} \text{.}
\end{align}
Using  (\ref{tcorr}) and (\ref{ordering-Bplasma}) to order the LHS, and (\ref{xyz-order}) to order the partial derivatives on the RHS of (\ref{induction}), we obtain an ordering for the induced electric field components
\begin{align} \label{eq:ordering-Einduced}
E_z^p \sim \delta^2 E^p \sim \delta \alpha E^p \ll E_y^p \sim \delta E^p \ll E_x^p \sim E^p \sim \delta v_{\text{t,i}} B^p \text{.}
\end{align}
In order to neglect $E_x^p$ and $E_y^p$ compared to their electrostatic counterparts we require $\delta B^{p} \ll B^{c}$, which is automatically satisfied if (\ref{ordering-beta}) holds. It follows that $E_z^p$ can also be neglected, because $E_z^p \ll E_y^p$ (from (\ref{eq:ordering-Einduced})) and the neglect of $E_y^p$ has been justified. This discussion justifies the electrostatic approximation and hence the use of $\vec{E} = - \nabla \phi$ in the equations of this paper.

We note that our orderings do not preclude a larger parallel current $\vec{j}^L$ (e.g. due to divertor target potential bias, Edge Localized Mode disruptions \cite{Kirk-2006} etc.) with $j_x^L = - j^L \sin\alpha $, $j^L_y = 0$ and $j_z^L = j^L \cos \alpha$, provided that the magnetic field produced by the plasma in the magnetic presheath remains much smaller than the external one and that the electrostatic assumption remains valid. Such a current would have to satisfy $\nabla \cdot \vec{j}^L = 0$ independently. In \ref{appendix:largejz}, we show that our equations allow for a parallel current density of size $j^L \ll \left( \alpha/\beta \right) \delta en_{\text{i}}v_{\text{t,i}}$. This current density can be large, $j^L \gg j_z^D \sim \delta n_{\text{i}} e v_{\text{t,i}}$, because $\alpha \sim 0.1 \gg \beta \sim 0.004$ in the magnetic presheath.

\section{Single particle motion in system with $\alpha = 0$ and $\delta = 0$}

In this section we introduce the transformation to an alternative set of variables which we use to describe the particle motion in the field geometry of Figure \ref{figure-boundary-layers}. We consider the magnetic field exactly perpendicular to the electric field, equivalent to setting $\alpha = 0$, and we also take $\delta = 0$. We therefore start by analyzing the problem of a single charged particle moving in a strongly varying electric field directed in the $x$ direction and a constant magnetic field directed in the $z$ direction, which we refer to as the zeroth order problem. This is introduced in Section 3.1.  

The phase space coordinates of the particle are given by $\boldsymbol{\xi} = (\vec{r}, \vec{v})$, where $\vec{r} = (x,y,z)$ is the position vector and $\vec{v} = d\vec{r} / dt =  (v_x, v_y, v_z)$ is the velocity vector of the particle. The zeroth order problem is not analytically solvable for a general strongly varying electric field, but it does have physically meaningful constants of integration: total energy $U$, perpendicular energy $U_{\perp}$, and orbit position $\bar{x}$, which we calculate in Section 3.1. Knowledge of these zeroth order constant parameters allows us to obtain $v_z$, gives a set of allowed values for $x$ and allows $v_x$ and $v_y$ to be determined from $x$. Hence, the constants of integration will be called orbit parameters. 
After identifying the conditions for this system to exhibit periodicity, in Section 3.2 we define the gyrophase $\varphi$ of the zeroth order problem. In addition to the orbit parameters, knowledge of $\varphi$ allows us to identify the exact position $x$, and therefore the exact velocity components $v_y$ and $v_x$.

The new orbit parameters and the phase $\varphi$ constitute a full set of alternative coordinates to describe the zeroth order problem. In this problem, $y$ and $z$ are symmetry directions and do not matter because nothing depends on them. In Section 3.3 we introduce the change of variables to this alternative set of coordinates. If we allow $\alpha \ll 1$ and $\delta \ll 1 $, we expect the orbit parameters to vary significantly over a timescale much longer than the orbital one, specifically the magnetic presheath timescale $t_{\text{MPS}}$ of (\ref{tMPS}). In Section 3.4 we introduce an adiabatic invariant $\mu$ for the grazing angle magnetic presheath problem \cite{Cohen-Ryutov-1998}, which is approximately conserved for $\alpha \ll 1$ and $\delta \ll 1$ even over timescales comparable to $t_{\text{MPS}}$. It is a generalization of the magnetic moment to the problem under consideration, and can be used instead of $U_{\perp}$ to provide a new complete set of coordinates for the particle motion.

\subsection{Orbit parameters}

We consider single particle motion in the system where $\alpha = \delta = 0$ exactly, 
with equations of motion (EOMs)
\begin{align} \label{x-EOM}
\dot{x} = v_x \text{,}
\end{align}
\begin{align} \label{y-EOM}
\dot{y} = v_y \text{,}
\end{align}
\begin{align} \label{z-EOM}
\dot{z} = v_z \text{,}
\end{align}
\begin{align} \label{vx-EOM}
\dot{v}_x = -\frac{Ze}{m_{\text{i}}}\frac{d \phi (x)}{d x} + \Omega v_{y} \text{,}
\end{align}
\begin{align} \label{vy-EOM}
\dot{v}_y = - \Omega v_{x} \text{,}
\end{align}
\begin{align} \label{vz-EOM}
\dot{v}_z = 0 \text{,}
\end{align}
where an overdot $\dot{}$ denotes taking a time derivative $d/dt$. In this section we derive some conserved quantities, or orbit parameters, from these equations.

Equation (\ref{vy-EOM}) is integrated to 
\begin{align} \label{vy-xbar-def}
\bar{x} = \frac{1}{\Omega}v_y + x \sim \rho_{\text{i}}  \text{,}
\end{align}
where the constant of integration $\bar{x}$ is the value of the $x$ coordinate for which $v_y$ is zero. This quantity represents the position of the orbit as a whole in the $x$ direction, and is ordered $\rho_{\text{i}}$ like the typical scale lengths in this direction in the magnetic presheath.

Multiplying the EOMs (\ref{vx-EOM}) and (\ref{vy-EOM}) by the respective velocity component and adding them results in conservation of perpendicular energy, $
\dot{U}_{\perp}= 0 \text{,} $ with $U_{\perp}$ defined by
\begin{align} \label{perp-energy-definition}
U_{\perp} = \frac{1}{2} v_{x}^2 + \frac{1}{2} v_{y}^2 + \frac{Ze \phi (x)}{m_{\text{i}}} \sim v_{\text{t,i}}^2 \text{.}
\end{align}
Making use of the constants of motion $\bar{x}$ and $U_{\perp}$ and introducing an effective potential \cite{Cohen-Ryutov-1998, Holland-Fried-Morales-1993, Gerver-Parker-Theilhaber-1990},
\begin{align} \label{chi-def}
\chi\left(x, \bar{x} \right) = \frac{1}{2}\Omega ^2 \left( x - \bar{x} \right) ^2 + \frac{Ze \phi (x)}{m_{\text{i}}} \sim v_{\text{t,i}}^2  \text{,}
\end{align}
we can express $v_{x}$ as 
\begin{align} \label{vx}
v_{x} = \sigma_x \sqrt{2\left(U_{\perp} - \chi \left( x, \bar{x} \right) \right)} \text{,}
\end{align}
where $\sigma_x = \pm 1$ depending on whether the particle is moving towards the top or bottom of the orbit respectively (when $x$ is seen as the vertical direction).

We introduce a further constant of motion, obtained by multiplying (\ref{vz-EOM}) by $v_z$, integrating in time and adding the result to the perpendicular energy. This constant of the motion is the total energy of the ion,
\begin{align} \label{U-def}
U = \frac{1}{2} v_{x}^2 + \frac{1}{2} v_{y}^2 + \frac{1}{2}v_{z}^2 + \frac{Ze \phi (x)}{m_{\text{i}}} \sim v_{\text{t,i}}^2 \text{.}
\end{align}

\subsection{Periodicity, gyrophase and the gyroaverage}

\begin{figure}
\centering
\includegraphics[width = 0.5\textwidth]{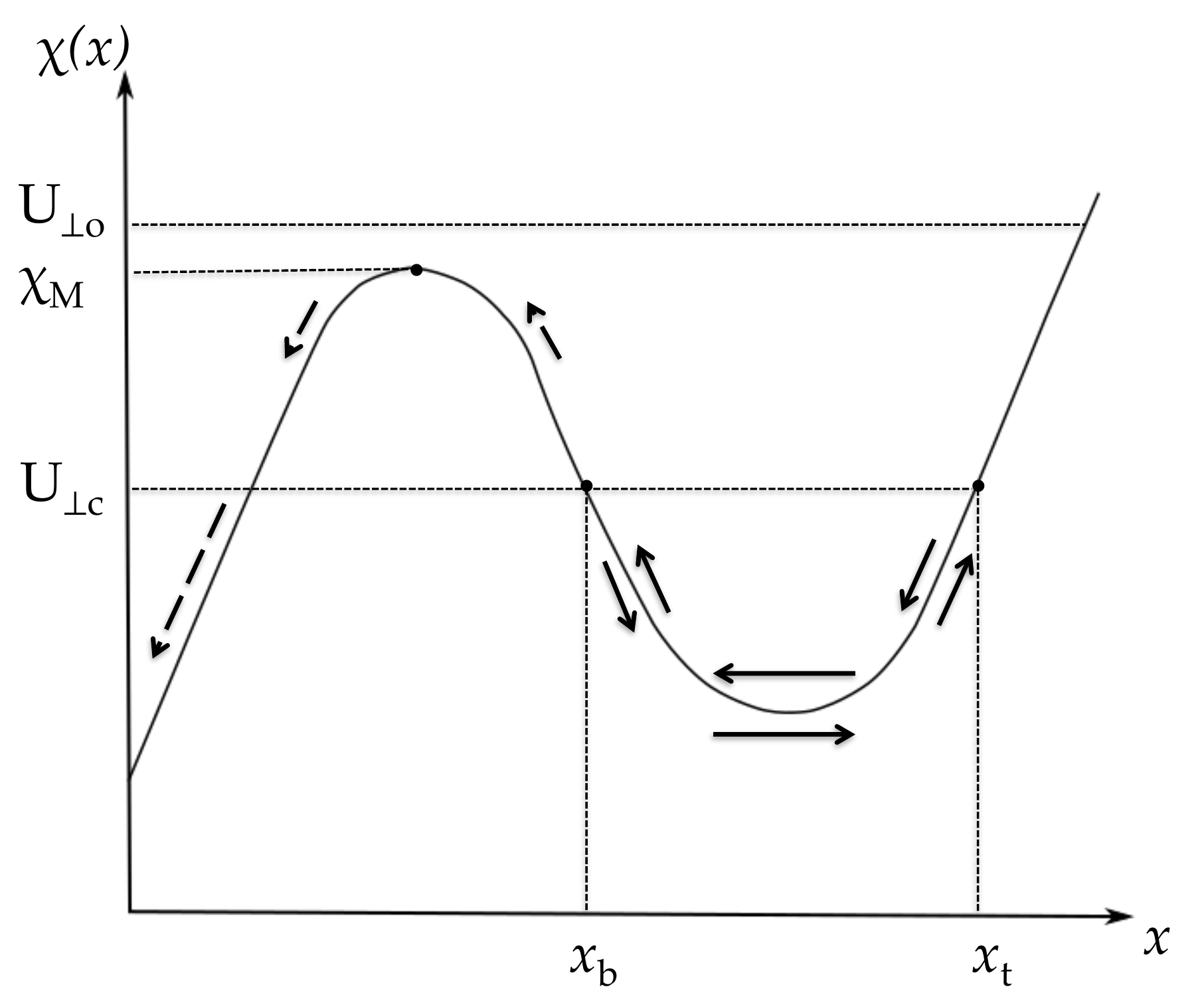}
\caption{A possible form for the effective potential curve $\chi \left( x, \bar{x} \right)$, determined by the value of $\bar{x}$ and the functional form of $\phi (x)$. If there is a local minimum for such a curve, the particle can be trapped within it (full arrows, energy $U_{\perp \text{c}}$). The bottom ($x_{\text{b}}$) and top ($x_{\text{t}}$) of such a closed orbit are labeled. If there is a local maximum $\chi_{\text{M}}$ and the perpendicular energy exceeds it (dashed arrows, energy $U_{\perp \text{o}}$), the particle is in an open orbit which intersects the wall.}
\label{figure-eff-pot}
\end{figure}

Having found the constant orbit parameters, we proceed to identify the phase of the orbit. Phase can only be used to describe motion that is periodic. The condition for the zeroth order motion to be periodic is that the particle be trapped around a minimum of the effective potential $\chi$ in (\ref{chi-def}). This is evident from (\ref{vx}) by requiring that $v_x$ always be real, and from Figure \ref{figure-eff-pot}. The bounce points $x_{\text{b}}$ and $x_{\text{t}}$ are the bottom and top of the orbit respectively, and they are obtained by setting equation (\ref{vx}) to zero for some combination of $U_{\perp}$ and $\bar{x}$. 

With the particle exhibiting periodic motion, the zeroth order generalized gyrofrequency $\overline{\Omega}$, which is ordered $\sim \Omega$, is given by
\begin{align} \label{Omegabar-def}
\frac{2\pi}{\overline{\Omega}} = 2 \int_{x_{\text{b}}}^{x_{\text{t}}} \frac{dx}{\left|v_x \right|} = 2 \int_{x_{\text{b}}}^{x_{\text{t}}} \frac{dx}{\sqrt{2\left(U_{\perp} - \chi \left(x, \bar{x}\right) \right)}} \text{,}
\end{align}
where the factor of $2$ is due to the particle going from $x_{\text{b}}$ to $x_{\text{t}}$ and then coming back to $x_{\text{b}}$.
We define the zeroth order gyrophase $\varphi = \overline{\Omega} t $ using $dt = dx / v_x$ and equation (\ref{vx}) for $v_x$. Setting $\varphi = 0$ at $x = x_{\text{t}}$, we have
\begin{align} \label{varphi-def}
\varphi = \sigma_x \overline{\Omega}\int_{x_{\text{t}}}^x  \frac{1}{\sqrt{2\left(U_{\perp} - \chi \left(x', \bar{x}\right) \right)}} dx^{\prime} \text{.}
\end{align}
This equation gives $\varphi (x, U_{\perp}, \bar{x}, \sigma_x)$, a quantity between $-\pi$ and $\pi$. 

The gyroaverage is defined as an average of some quantity over a complete ion orbit,
\begin{align} \label{gyroav-def}
\left\langle \ldots \right\rangle_{\varphi} \equiv \frac{1}{2\pi} \oint \left( \ldots \right) d\varphi = \frac{\overline{\Omega}}{\pi} \int_{x_{\text{b}}}^{x_{\text{t}}} \frac{\left( \ldots \right)}{\sqrt{2\left(U_{\perp} - \chi \left(x, \bar{x} \right)\right)}} dx \text{,}
\end{align}
which is performed holding all the orbit parameters fixed. Note that the operation in (\ref{gyroav-def}) is performed holding also $y$ and $z$ fixed, which is uninteresting for the moment because they are symmetry directions in the zeroth order problem, but will become relevant later. For practical purposes, gyroaverages are best taken using the second integral in $x$, which is written (for simplicity) assuming that the quantity being gyroaveraged is independent of $\sigma_x$, equivalent to assuming that it is symmetric about $\varphi = 0$ (which is valid for all gyroaveraged quantities in this paper).

\subsection{Change of variables}

We have shown that, given the set of coordinates $\boldsymbol{\xi}$, we can define the constants $\bar{x}$ from $x$ and $v_y$ using (\ref{vy-xbar-def}), $U_{\perp}$ from $x$, $v_x$ and $v_y$ using (\ref{perp-energy-definition}), and $U$ from $x$, $v_x$, $v_y$ and $v_z$ using (\ref{U-def}). Exploiting those constants, we can define the gyrophase $\varphi$ in a less trivial way from the constants $U_{\perp}$, $\bar{x}$ and the variable $x$ using (\ref{varphi-def}). The only variables of the orbital motion that remain to be defined are to do with the $y$ and $z$ coordinates of the gyrating particle. These are symmetry directions in the zeroth order problem, therefore we keep the $y$ and $z$ coordinates unchanged.

There is a full set of alternative coordinates to describe the particle motion in the zeroth order problem, so we introduce the change of variables
\begin{align} \label{xi-eta0}
\boldsymbol{\xi} = \left(\vec{r}, \vec{v}\right) \rightarrow \boldsymbol{\eta} = \left(\bar{x}, y, z, U_{\perp}, \varphi, U, \sigma_{\parallel} \right) \text{.}
\end{align}
Note that there are seven quantities instead of six in the vector $\boldsymbol{\eta}$, because we introduced the discrete variable $\sigma_{\parallel} = \pm 1$ in order to allow for positive and negative $v_z$ (see equation (\ref{vz-U-Uperp}) below).
We denote the inverse of (\ref{varphi-def}), which gives $x$ as a function of $\varphi$, $\bar{x}$ and $U_{\perp}$, as $x_{\text{gk}}\left( \bar{x}, U_{\perp}, \varphi \right)$ or $x_{\varphi}$,
\begin{align} \label{xgk-def}
x = x_{\text{gk}}(\bar{x}, U_{\perp}, \varphi) \equiv  x_{\varphi} \text{.}
\end{align}
Differentiating (\ref{xgk-def}) with respect to time we get, by applying the chain rule and using $\dot{\varphi} = \overline{\Omega}$, $\dot{\bar{x}} = 0$ and $\dot{U}_{\perp} = 0$,
\begin{align} \label{vx-f-gyro}
v_x  = \overline{\Omega}\frac{\partial x_{\text{gk}}}{\partial \varphi} \left(\bar{x}, U_{\perp}, \varphi\right) \text{.}
\end{align}

Using (\ref{vy-xbar-def}), we can express $v_y$ as a function of the same three coordinates. However, it turns out to be more useful to express $v_y$ as a function of $\bar{x}$, $U_{\perp}$ and $\varphi$ by using the time derivative of (\ref{vx-f-gyro}),
\begin{align}
\frac{d v_x}{dt} = \overline{\Omega}^2 \frac{\partial^2 x_{\text{gk}}}{\partial \varphi^2}  \text{,}
\end{align}
and (\ref{vx-EOM}), to get
\begin{align} \label{vy-reexpressed}
v_y = \frac{\overline{\Omega}^2}{\Omega}\frac{\partial^2 x_{\text{gk}}}{\partial \varphi^2} \left(\bar{x}, U_{\perp}, \varphi\right) + \frac{1}{B} \frac{d \phi}{d x}\left( x_{\text{gk}} \left(\bar{x}, U_{\perp}, \varphi\right) \right) \text{ .}
\end{align}
From (\ref{perp-energy-definition}) and (\ref{U-def}), $v_z$ is a simple function of $U_{\perp}$, $U$ and $\sigma_{\parallel}$,
\begin{align} \label{vz-U-Uperp}
v_z = v_{\parallel}\left(U_{\perp}, U, \sigma_{\parallel} \right) = \sigma_{\parallel} \sqrt{2\left( U - U_{\perp} \right)} \text{,}
\end{align}
where $\sigma_{\parallel}$ distinguishes between particles moving in the $+z$ direction, with $\sigma_{\parallel}  = 1$, and ones moving in the $-z$ direction, with $\sigma_{\parallel} = - 1$.

\subsection{The adiabatic invariant}

Periodic solutions of Hamiltonian systems always have an adiabatic invariant associated with their motion: a quantity that is conserved even when the system is slowly perturbed (i.e. parameters of the original system are changed slowly). The zeroth order system allows for closed (periodic) orbits, so we expect an adiabatic invariant to be present. We calculate it in \ref{appendix:mucalc} and show that it is given by 
\begin{align} \label{mu-gyroav}
\mu = \frac{1}{\overline{\Omega}} \left\langle v_{x}^2 \right\rangle_{\varphi} \text{.}
\end{align}
This can be re-expressed as a function $\mu_{\text{gk}}$, which is the definite integral
\begin{align} \label{mu(Uperp,xbar)}
\mu = \mu_{\text{gk}} \left( \bar{x}, U_{\perp}\right) = \frac{1}{\pi} \int_{x_{\text{b}}}^{x_{\text{t}}} \sqrt{2\left( U_{\perp} - \chi\left(x, \bar{x}\right) \right) } dx \sim  v_{\text{t,i}} \rho_{\text{i}}  \text{.}
\end{align}

Equation (\ref{mu(Uperp,xbar)}) is a generalization of the usual magnetic moment to the grazing angle presheath geometry which we study in this paper. It was derived by Cohen and Ryutov in reference \cite{Cohen-Ryutov-1998}.
From (\ref{mu-gyroav}), it is easy to check that the adiabatic invariant recovers the usual magnetic moment for a weakly varying electric field. Note that we can use $\mu$ as an orbit parameter instead of $U_{\perp}$.

In \ref{appendix:linE}, we use the equations derived so far to solve explicitly the zeroth order problem with a linearly varying electric field in terms of the orbit parameters and gyrophase. This allows us to study the simplest case of orbit distortion, equivalent to changing the shape of the orbit from a circle to an ellipse in the frame where motion is periodic.

\section{The gyrokinetic equation for a collection of ions in the magnetic presheath}

In this section, we extend the study of motion of a single ion in the system with $\alpha = 0$ and $\delta = 0$ described in Section 3 to a collection of ions moving in the system with $\alpha \ll 1$ and $\delta \ll 1$, which models a grazing angle magnetic presheath. In order to do this, we begin Section 4.1 by changing variables in the kinetic equation. We find that the distribution function has no lowest order dependence on gyrophase. We exploit this to derive the gyrokinetic equation (the gyroaverage of the kinetic equation) to next order, which encodes the variation of the distribution function over the typical timescale $t_{\text{MPS}}$ it takes for an ion to cross the magnetic presheath and reach the wall. In Section 4.2, we calculate the gyroaveraged time derivatives of the orbit parameters, which appear in the gyrokinetic equation, by analysing the motion of a single particle.

In Section 4.3 we return to the adiabatic invariant $\mu$ and explicitly show its invariance with the results of this section. We then demonstrate, in Section 4.4, the existence of another constant of the motion, which is conserved to first order in $\alpha$ and $\delta$. This constant, $y_{\star}$, is proportional to the canonical momentum in the $z$ direction \cite{Holland-Fried-Morales-1993}. We introduce a coordinate transformation in which $y$ and $U_{\perp}$, which are quantities that vary to first order, are replaced by $y_{\star}$ and $\mu$, both constants of the motion to first order. The new set of variables is denoted $\boldsymbol{\eta}_{\star}$. 

Finally, in Section 4.5, we summarize the results of Sections 4.1 and 4.2 by writing the most general gyrokinetic equation using the set of variables $\boldsymbol{\eta}$ (introduced in Section 3), outlining the boundary conditions one would need to impose in order to solve it. We then exploit the set of variables $\boldsymbol{\eta}_{\star}$ introduced in Section 4.4 to simplify the gyrokinetic equation, and solve it in the limit of a collisionless magnetic presheath $\rho_{\text{i}} \ll \alpha \lambda_{\text{mfp}}$. The solution of this gyrokinetic equation relies on knowledge of the magnetic presheath entrance distribution function.

\subsection{Change of variables in the kinetic equation}

We consider the behaviour of a system of ions in the magnetic presheath, where electric fields perpendicular to the wall are strong and strongly varying. The Vlasov equation, with an ion distribution function $f = f \left( \vec{r}, \vec{v}, t \right) = f \left( \boldsymbol{\xi}, t \right) $ and a general collision term $C\left[ f \right]$, is
\begin{align}
\frac{\partial f}{\partial t} + \vec{v} \cdot \frac{\partial f}{\partial \vec{r}} + \dot{\vec{v}} \cdot \frac{\partial f}{\partial \vec{v}} = C\left[ f\right] \text{.}
\end{align}

With our knowledge of the zeroth order single particle problem, we change variables to $\boldsymbol{\eta} = \left( \bar{x}, y, z, U_{\perp}, \varphi , U , \sigma_{\parallel} \right)$ with the aim to gyroaverage the kinetic equation. The new variables are meaningful because the motion of a single particle in the magnetic presheath is, to zeroth order, the orbit that we have solved for in Section 3, with a slow variation of the orbit parameters due to non-zero but small $\alpha$ and $\delta$.
With the new variables, the distribution function has a different form, which we denote
$F = F \left( \bar{x}, y, z, U_{\perp}, \varphi, U, \sigma_{\parallel}, t \right) = F \left( \boldsymbol{\eta}, t \right)$. The collision operator is re-expressed as $\mathcal{C}\left[F\right]$. Then, we write the kinetic equation
\begin{align} \label{kinetic-gyro}
\frac{\partial F}{\partial t} + \dot{\bar{x}} \frac{\partial F}{\partial \bar{x}} + \dot{y}\frac{\partial F}{\partial y} + \dot{z}\frac{\partial F}{\partial z} + \dot{U}_{\perp} \frac{\partial F}{\partial U_{\perp}} + \dot{U} \frac{\partial F}{\partial U} + \dot{\varphi} \frac{\partial F}{\partial \varphi} = \mathcal{C}[F] \text{.}
\end{align}
We expand the distribution function in pieces of order $F$, $\alpha F$, etc.,
\begin{align}
F = F_0 + F_1 + \ldots \text{.}
\end{align}
We point out that at $x \rightarrow \infty$ a major simplification occurs which stems from the fact that the electric field and its variation over an orbit become smaller (dominated by the turbulent electric fields), $\partial \phi / \partial x \sim \delta T/ \rho_{\text{i}} e$ and $\partial^2 \phi / \partial x^2 \sim \left( \delta / \rho_{i} \right) \partial \phi / \partial x$. This results in the familiar low-drift gyrokinetic circular orbits with magnetic moment given by $\mu = v_{\perp}^2/2\Omega$ with $v_{\perp}^2 = 2\left( U_{\perp} - Ze\phi /m_{\text{i}} \right)$, and gyrophase given by $\tan \varphi = v_x / v_y$.

We assume that the collision frequency is much smaller than the gyrofrequency, equivalent to assuming $\rho_{\text{i}} \ll \lambda_{\text{mfp}}$, and we order $\mathcal{C}[F_0] = O(\alpha \Omega F)$. Taking $O(\Omega F)$ terms only in (\ref{kinetic-gyro}) leads to the lowest order equation 
\begin{align} \label{gyro1}
\overline{\Omega} \frac{\partial F_0}{\partial \varphi} = 0 \text{.}
\end{align}
To obtain this, we have used the ordering for the time dependence in (\ref{tcorr}), and that $\dot{\varphi} = \overline{\Omega} \sim \Omega$, $\dot{\bar{x}} / \bar{x} \sim \dot{y} / y  \sim \dot{U}_{\perp} / U_{\perp} \sim \alpha \Omega $, and $\dot{z} / z \sim \dot{U} / U \sim \alpha^2 $. These orderings are a consequence of the orbit parameters being constants of motion in the zeroth order problem where $\alpha = \delta = 0$, and of $\dot{y} = v_y \sim v_{\text{t,i}} \sim \alpha \Omega l$ and $\dot{z} = v_z \sim v_{\text{t,i}} \sim \alpha^2 \Omega \left( l/\alpha \right) $. The variation of $U$ is second order because
\begin{align} \label{Udot1}
\dot{U} = \frac{Ze}{m_{\text{i}}} \frac{\partial \phi}{\partial t} = O\left( \alpha^2 \Omega v_{\text{t,i}}^2 \right) \simeq 0 \text{.}
\end{align}
This can be explicitly shown from the exact EOMs (\ref{vx-EOM-exact})-(\ref{vz-EOM-exact}) in the next subsection, but it is just the statement that the total energy of a particle is conserved up to explicit time dependence of the potential, and such time dependence is second order due to (\ref{tcorr}).

From (\ref{gyro1}), we deduce that the lowest order distribution function is gyrophase independent.
Then, the first order of (\ref{kinetic-gyro}) is
\begin{align} \label{kinetic-gyro-1}
\dot{\bar{x}} \frac{\partial F_0}{\partial \bar{x}} + \dot{y}\frac{\partial F_0}{\partial y} + \dot{U}_{\perp} \frac{\partial F_0}{\partial U_{\perp}} + \dot{\varphi} \frac{\partial F_1}{\partial \varphi} = \mathcal{C}[F_0]  \text{,}
\end{align}
where we neglected the second order $t$, $z$ and $U$ dependence. 
We take the gyroaverage of (\ref{kinetic-gyro-1}) to obtain the gyrokinetic equation,
\begin{align} \label{gyrokinetic-1}
\left\langle \dot{\bar{x}} \right\rangle_{\varphi} \frac{\partial F_0}{\partial \bar{x}} + \left\langle \dot{y} \right\rangle_{\varphi} \frac{\partial F_0}{\partial y}  + \left\langle \dot{U}_{\perp} \right\rangle_{\varphi} \frac{\partial F_0}{\partial U_{\perp}} = \left\langle \mathcal{C}[F_0] \right\rangle_{\varphi}  \text{.}
\end{align}
In the next section, we calculate the gyroaveraged time derivatives that appear in (\ref{gyrokinetic-1}).

\subsection{Calculating the time derivatives that appear in the kinetic equation}

The exact equations of motion (EOMs) for an ion moving in the collisionless magnetic presheath of Figure \ref{figure-boundary-layers}, with the set of axes shown there, are (\ref{x-EOM})-(\ref{z-EOM}) and
\begin{align}
\label{vx-EOM-exact}
\dot{v}_x = -\frac{Ze}{m_{\text{i}}}\frac{\partial \phi}{\partial x}\left(x, y, z, t\right) + \Omega v_{y}\cos\alpha \text{,}
\end{align}
\begin{align}
\label{vy-EOM-exact}
\dot{v}_y = -\frac{Ze}{m_{\text{i}}}\frac{\partial \phi}{\partial y}\left(x, y, z, t\right) - \Omega v_{x}\cos\alpha - \Omega v_{z}\sin\alpha \text{,}
\end{align}
\begin{align}\label{vz-EOM-exact}
\dot{v}_z = -\frac{Ze}{m_{\text{i}}}\frac{\partial \phi}{\partial z}\left(x, y, z, t\right) + \Omega v_{y}\sin\alpha \text{.}
\end{align}
Note that expanding all the terms on the RHS of (\ref{vx-EOM-exact})-(\ref{vz-EOM-exact}) in $\alpha$ and $\delta$ and retaining only lowest order ones recovers (\ref{vx-EOM})-(\ref{vz-EOM}). In this section we will exploit the equations we derived in Section 3, which need to be updated to include $y$, $z$, and $t$ dependence due to $\phi$, but have the same form. From here on we omit the weak $z$ and $t$ dependence of the potential as it leads to second order effects, not treated here. 

Differentiating (\ref{vy-xbar-def}) gives 
\begin{align} \label{dxbardt}
\dot{\bar{x}} = \frac{1}{\Omega} \dot{v}_y + v_x \text{.}
\end{align}
Inserting (\ref{vy-EOM-exact}) into (\ref{dxbardt}) we get, to first order,
\begin{align} \label{dxbardt2}
\dot{\bar{x}} = -\alpha v_{\parallel}\left(U_{\perp}, U, \sigma_{\parallel}\right) - \frac{1}{B}\frac{\partial \phi}{\partial y}\left(x_{\varphi}, y \right) + O\left(\alpha^2 \Omega \rho_{\text{i}}\right) \text{,}
\end{align}
where $x_{\varphi}$ is defined in (\ref{xgk-def}) and $v_{\parallel}$ is defined in (\ref{vz-U-Uperp}).
This leads to
\begin{align} \label{xbardot}
\left\langle \dot{\bar{x}}\right\rangle_{\varphi} = -\alpha v_{\parallel}\left(U_{\perp}, U, \sigma_{\parallel} \right) - \frac{1}{B} \left\langle \frac{\partial \phi}{\partial y}\left(x_{\varphi}, y\right) \right\rangle_{\varphi} + O\left(\alpha^2 \Omega \rho_{\text{i}}\right) \text{.}
\end{align}
This is the velocity at which the orbit as a whole moves in the $x$ direction, normal to the wall, and has two contributions. The first term on the RHS is from the particle motion parallel to the magnetic field, whose component perpendicular to the wall is small. The second one is from the $\vec{E} \times \vec{B}$ drift normal to the wall, due to the small electric field in the $y$ direction. In Section 2 we had mentioned that these two effects were expected to be of the same order when $\delta \sim \alpha$.

In Section 3 we wrote expression (\ref{vy-reexpressed}) for $\dot{y} = v_y$ in terms of the variables $\boldsymbol{\eta}$. From it, we obtain by gyroaveraging
\begin{align} \label{ydot}
\left\langle \dot{y} \right\rangle_{\varphi} = \frac{1}{B}\left\langle \frac{\partial \phi}{\partial x} \left(x_{\varphi}, y\right) \right\rangle_{\varphi} + O(\alpha^2 \Omega l) \text{,}
\end{align}
where we have used that the gyroaverage of a gyrophase derivative is zero. Equation (\ref{ydot}) is the $\vec{E} \times \vec{B}$ drift in a strongly varying electric field (almost) perpendicular to a magnetic field.

Differentiating $U_{\perp}$ given in (\ref{perp-energy-definition}) with respect to time we have, in terms of the phase space variables $\boldsymbol{\xi}$,
\begin{align} \label{dUperpdt}
\dot{U}_{\perp} = v_x \dot{v}_x + v_y \dot{v}_y + \frac{Ze}{m_{\text{i}}}\dot{\phi}\left(x, y\right) \text{.}
\end{align}
The last term can be rewritten, by use of the chain rule, as
\begin{align}
\dot{\phi}\left(x, y \right) = v_x \frac{\partial \phi}{\partial x} \left( x, y \right) + v_y \frac{\partial \phi}{\partial y} \left(x, y \right) + O\left( \alpha^2 \Omega m_{\text{i}} v_{\text{t,i}}^2 / e \right) \text{,}
\end{align}
where the neglected terms are $v_z \partial \phi / \partial z$ and $\partial \phi / \partial t$.
This leads to
\begin{align} \label{dUperpdt}
\dot{U}_{\perp} = v_x \dot{v}_x + v_y \dot{v}_y + \frac{Ze}{m_{\text{i}}}v_x\frac{\partial \phi}{\partial x}\left(x, y\right) + \frac{Ze}{m_{\text{i}}}v_y\frac{\partial \phi}{\partial y}\left(x, y\right) + O\left(\alpha^2 \Omega v_{\text{t,i}}^2 \right) \text{.}
\end{align}
Substituting the EOMs (\ref{vx-EOM-exact}) and (\ref{vy-EOM-exact}) we obtain, after some cancellations, 
\begin{align}
\dot{U}_{\perp} = -\alpha \Omega v_z v_y + O\left(\alpha^2 \Omega v_{\text{t,i}}^2 \right) \text{,}
\end{align}
which can be expressed as a function of gyrokinetic variables using (\ref{vy-reexpressed}) and (\ref{vz-U-Uperp}),
\begin{align} \label{dUperpdt2}
\dot{U}_{\perp} = - \alpha \Omega v_{\parallel} \left( U_{\perp}, U, \sigma_{\parallel} \right) \left( \frac{\overline{\Omega}^2}{\Omega}\frac{\partial^2 x_{\varphi}}{\partial \varphi^2} + \frac{1}{B} \frac{\partial \phi}{\partial x}\left(x_{\varphi}, y\right) \right) + O\left(\alpha^2 \Omega v_{\text{t,i}}^2\right) \text{.}
\end{align} 
The gyroaveraged time derivative of $U_{\perp}$ is therefore
\begin{align} \label{Uperpdot}
\left\langle \dot{U}_{\perp} \right\rangle_{\varphi} = - \alpha \Omega v_{\parallel} \left( U_{\perp}, U, \sigma_{\parallel} \right) \frac{1}{B} \left\langle \frac{\partial \phi}{\partial x}\left(x_{\varphi}, y\right) \right\rangle_{\varphi} + O\left(\alpha^2 \Omega v_{\text{t,i}}^2\right) \text{.}
\end{align}

\subsection{Adiabatic invariance to first order}

Having obtained the gyroaveraged time derivatives in (\ref{gyrokinetic-1}), we verify that the adiabatic invariant is conserved to first order. This is important because the adiabatic invariant $\mu$ may be used as an alternative variable to the perpendicular energy $U_{\perp}$.

From (\ref{mu(Uperp,xbar)}) updated to include $y$ dependence, the adiabatic invariant is a function of the new variables, $\mu = \mu_{\text{gk}} \left(\bar{x},y, U_{\perp} \right)$. By differentiating this function we obtain, as shown in \ref{appendix:muderivatives},
\begin{align} \label{dmudxbar}
\frac{\partial \mu}{\partial \bar{x}} = - \frac{Ze}{m_{\text{i}}\overline{\Omega}} \left\langle \frac{\partial \phi}{\partial x}\left(x_{\varphi}, y\right) \right\rangle_{\varphi} \text{,}
\end{align}
\begin{align} \label{dmudY}
\frac{\partial \mu}{\partial y} = -\frac{Ze}{m_{\text{i}}\overline{\Omega}} \left\langle \frac{\partial \phi}{\partial y}\left(x_{\varphi}, y\right) \right\rangle_{\varphi}\text{,}
\end{align}
\begin{align} \label{dmudUperp}
\frac{\partial \mu}{\partial U_{\perp}} = \frac{1}{\overline{\Omega}} \text{.}
\end{align} 
  
Using the chain rule to take the time derivative $\dot{\mu}$ and gyroaveraging, the first order gyroaveraged total derivative of the magnetic moment with respect to time is 
\begin{align}
\left\langle \dot{\mu} \right\rangle_{\varphi} = \frac{\partial \mu}{\partial \bar{x}} \left\langle \dot{\bar{x}} \right\rangle_{\varphi} + \frac{\partial \mu}{\partial y} \left\langle \dot{y} \right\rangle_{\varphi}  + \frac{\partial \mu}{\partial U_{\perp}} \left\langle \dot{U}_{\perp} \right\rangle_{\varphi} \text{.}
\end{align}
Upon using (\ref{dmudxbar})-(\ref{dmudUperp}) and (\ref{xbardot}), (\ref{ydot}) and (\ref{Uperpdot}) we obtain adiabatic invariance to first order, 
\begin{align} \label{ad-invariance}
\left\langle \dot{\mu} \right\rangle_{\varphi} = O(\alpha^2 \Omega \mu ) \simeq 0 \text{.}
\end{align} 
Note that (\ref{dmudxbar})-(\ref{dmudUperp}) provide alternative ways to express the gyrofrequency and the gyroaveraged $x$ and $y$ components of the electric field, which appear in (\ref{xbardot}), (\ref{ydot}) and (\ref{Uperpdot}), in terms of partial derivatives of $\mu$ \cite{Cohen-Ryutov-1998}.

\subsection{A first order constant of motion and an alternative transformation of variables}

In Section 3, we obtained constants of integration of the zeroth order problem, which were the orbit parameters $\bar{x}$, $U_{\perp}$ and $U$. When $\alpha \sim \delta \ll 1 $, we found that $\bar{x}$ and $U_{\perp}$ vary to first order in $\alpha$, so that $\dot{\bar{x}}/ \bar{x} \sim \dot{U}_{\perp}/ U_{\perp} \sim \alpha \Omega$, while $U$ varies to second order only due to the weak explicit dependence of the potential on time, $\dot{U} / U \sim \alpha^2 \Omega$. We have also found that the variation of $y$ with time is first order in our ordering, $\dot{y}/y \sim \alpha \Omega$, while the variation of $z$ is second order, $\dot{z}/z \sim \alpha^2 \Omega$. In what follows, we introduce another orbit parameter $y_{\star} \sim l$ which, like total energy $U$, is constant to first order, $\dot{y}_{\star} / y_{\star} \sim \alpha^2 \Omega$.

The EOM in the $z$ direction is (\ref{vz-EOM-exact}), which written to first order in $\alpha$ becomes
\begin{align}
\dot{v}_z = \alpha \Omega v_y \text{,}
\end{align}
where we have neglected the second order term $\partial \phi / \partial z \sim \alpha^2 T / e \rho_{\text{i}}$. Integrating this in time and introducing the constant of integration $y_{\star}$ we get
\begin{align}
y_{\star} = y - \frac{v_z}{\alpha \Omega} \sim l \text{.}
\end{align}
This quantity is proportional to the canonical momentum in the $z$ direction \cite{Holland-Fried-Morales-1993}, if the magnetic vector potential is written such that it has no $z$ dependence. Such a choice for the vector potential is, for example,
\begin{align} \label{mag-vec-pot}
\vec{A} = \begin{pmatrix}
0 \\ xB\cos\alpha  \\ -yB \sin \alpha 
\end{pmatrix} \text{.}
\end{align}
This vector potential can be checked by calculating the magnetic field that corresponds to it, 
\begin{align}
\vec{B} = \nabla \times \vec{A} = \begin{pmatrix}
-B \sin \alpha  \\ 0 \\ B\cos\alpha
\end{pmatrix} \text{,}
\end{align}
which is exactly the magnetic field present in the magnetic presheath. Using $\sin \alpha \simeq \alpha$, the canonical momentum in the $z$ direction, $p_z$, is proportional to $y_{\star}$, $p_z = m_{\text{i}} v_z + ZeA_z = m_{\text{i}} \left( v_z -  \Omega y \sin\alpha \right) \simeq -m_{\text{i}} \alpha \Omega y_{\star} $. Because the magnetic vector potential is written such that it is independent of $z$ and the electrostatic potential depends on $z$ only to second order, the canonical momentum that we have just calculated is a constant of the motion to first order, $\dot{p}_z / p_z \sim \alpha^2 \Omega$. Note that the orbit position $\bar{x}$ that we have calculated in Section 3 is proportional to the canonical momentum in the $y$ direction \cite{Gerver-Parker-Theilhaber-1990}, $p_y = m_{\text{i}} v_y + ZeA_y = m_{\text{i}} \left( v_y +  \Omega x \cos\alpha \right) \simeq m_{\text{i}} \Omega \bar{x} $. Because both the magnetic vector potential in (\ref{mag-vec-pot}) and the electrostatic potential have a first order dependence on $y$, we have $\dot{p}_y / p_y \sim \alpha \Omega$ as expected.

Using (\ref{vz-U-Uperp}) we express $y_{\star}$ in terms of variables in $\boldsymbol{\eta}$,
\begin{align} \label{ystar-y-vpar}
y_{\star} = y - \frac{1}{\alpha \Omega}v_{\parallel}\left(U_{\perp}, U, \sigma_{\parallel} \right) \text{,}
\end{align}
and introduce the coordinate transformation 
\begin{align} \label{eta-etastar}
\boldsymbol{\eta} = \left( \bar{x}, y, z, U_{\perp}, \varphi, U, \sigma_{\parallel}  \right) \rightarrow \boldsymbol{\eta}_{\star} = \left( \bar{x}, y_{\star}, z, \mu, \varphi, U, \sigma_{\parallel} \right) \text{.}
\end{align}
The fact that $y_{\star}$ is constant to first order implies that its gyroaveraged time derivative (while holding all other variables in $\boldsymbol{\eta}_{\star}$ fixed) is also zero to that order, 
\begin{align} \label{ystardot}
\left\langle \dot{y}_{\star} \right\rangle_{\varphi} = O\left( \alpha^2 \Omega y_{\star} \right) \simeq 0 \text{.}
\end{align} 
We show this in the context of the new variables. The gyroaveraged time derivative of $v_{\parallel}$ is, using the chain rule, (\ref{vz-U-Uperp}), (\ref{Udot1}) and (\ref{Uperpdot}),
\begin{align} \label{vparalleldot}
\left\langle \dot{v}_{\parallel} \left( U_{\perp}, U, \sigma_{\parallel} \right) \right\rangle_{\varphi} \simeq - \frac{1}{v_{\parallel}\left(U_{\perp}, U, \sigma_{\parallel} \right)} \left\langle \dot{U}_{\perp} \right\rangle_{\varphi}  = \frac{\alpha \Omega}{B} \left\langle \frac{\partial \phi}{\partial x}\left(x_{\varphi}, y\right) \right\rangle_{\varphi} + O(\alpha^2 \Omega v_{\text{t,i}}^2) \text{,}
\end{align}
which is $\alpha \Omega$ times the gyroaveraged time derivative of $y$ in (\ref{ydot}). It follows that the gyroaveraged time derivative of $y_{\star}$ is zero. %Note that here the gyroaveraging operation is performed holding the new variables $\boldsymbol{\eta}_{\star}$ fixed. 
In the next subsection, we exploit the change of variables (\ref{eta-etastar}) and the results (\ref{ad-invariance})  and (\ref{ystardot}) to write the gyrokinetic equation in a simpler form.

\subsection{Final gyrokinetic equations: in general form and simplified by taking the collisionless limit $\rho_{\text{i}} \ll \alpha \lambda_{\text{mfp}}$}

In this subsection, we write the most general gyrokinetic equation expressed in the set of variables $\boldsymbol{\eta}$ using the results of Sections 4.1 and 4.2. We also provide the boundary conditions that are necessary in order to solve the gyrokinetic equation in these variables. We then show that the gyrokinetic equation is simplified when expressed in terms of the set of variables $\boldsymbol{\eta}_{\star}$, and find the solution in the limit of a collisionless magnetic presheath, $ \rho_{\text{i}} \ll \alpha \lambda_{\text{mfp}}$.

A collection of ions moving in a grazing angle magnetic presheath can be described by a distribution function $F(\boldsymbol{\eta}, t) \simeq F_0 (\boldsymbol{\eta}, t)$ which must be gyrophase independent and must satisfy, to lowest order in $\alpha \sim \delta$ (using (\ref{gyrokinetic-1}), (\ref{xbardot}), (\ref{ydot}) and (\ref{Uperpdot})),
\begin{align} \label{gyrokinetic-full-explicit}
\left( -\alpha v_{\parallel}(U_{\perp}, U) - \frac{1}{B} \left\langle \frac{\partial \phi}{\partial y}(x_{\varphi}, y) \right\rangle_{\varphi} \right) \frac{\partial F_0}{\partial \bar{x}} + \frac{1}{B}\left\langle \frac{\partial \phi}{\partial x} \left(x_{\varphi}, y\right) \right\rangle_{\varphi}  \frac{\partial  F_0}{\partial y} \nonumber \\
- \alpha v_{\parallel}\frac{Ze}{m_{\text{i}}} \left\langle \frac{\partial \phi}{\partial x}\left(x_{\varphi}, y\right) \right\rangle_{\varphi} \frac{\partial F_0}{\partial U_{\perp}} = \left\langle \mathcal{C}[F_0] \right\rangle_{\varphi} \text{.}
\end{align}
This is the gyroaverage of the ion kinetic equation, or ion gyrokinetic equation. Note that in order to solve (\ref{gyrokinetic-full-explicit}) we require boundary conditions in terms of $\bar{x}$, $y$ and $U_{\perp}$. We assume that infinitely energetic particles ($U_{\perp} \rightarrow \infty$) are not allowed,
\begin{align} \label{bc-F0-Uperp}
\lim_{U_{\perp} \rightarrow \infty} F_0 = 0 \text{.}
\end{align}
Since $v_y = \Omega \left( \bar{x} - x \right) \sim \sqrt{U_{\perp}}$, this boundary condition implies that the limit $\bar{x} \rightarrow \infty$ is equivalent to the limit $x \rightarrow \infty$. The boundary condition for incoming ions at the magnetic presheath entrance $x \rightarrow \infty$ is therefore a boundary condition for $\bar{x}$, 
\begin{align} \label{bc-F0-xbar}
\lim_{\bar{x} \rightarrow \infty} F_0 = \lim_{x \rightarrow \infty} F_0 =
F_0^{\infty}\left( y, U_{\perp}, U, \sigma_{\parallel} \right) \text{ for } \left\langle \dot{\bar{x}} \right\rangle_{\varphi} < 0 \text{.}
\end{align}
Note that the boundary condition only gives the particles that are drifting into the presheath, $\left\langle \dot{\bar{x}} \right\rangle_{\varphi} < 0$.
The assumption of an electron repelling wall implies that no ion comes back from the wall. Thus, we need to use the following boundary condition for orbits that are sufficiently near the wall, 
\begin{align} \label{bc-F0-xbar-wall}
F_0 = 0 \text{ for \emph{open} orbits.}
\end{align}
Open orbits are ion orbits whose lowest order trajectory (calculated using the values of the orbit parameters at a given instant) intersects the wall. The range of values of $\bar{x}$ and $U_{\perp}$ for which an orbit is open is given in Section 5.1.

In the $y$ direction, we must impose that the ion density be zero at $y \rightarrow \pm \infty$ for incoming particles because this corresponds to being far away from the separatrix and, eventually, outside of the SOL (see Figures \ref{figure-combined-divertor} and \ref{diagram-SOL-thick}, and the discussion in \ref{appendix:turbulence}),
\begin{align} \label{bc-F0-y+}
\lim_{y \rightarrow \infty} F_0 = 0  \text{ for } \left\langle \dot{y} \right\rangle_{\varphi} < 0 \text{,}
\end{align}
\begin{align} \label{bc-F0-y}
\lim_{y \rightarrow -\infty} F_0 = 0  \text{ for } \left\langle \dot{y} \right\rangle_{\varphi} > 0 \text{.}
\end{align}

Applying the tranformation of variables (\ref{eta-etastar}) to the kinetic equation (\ref{kinetic-gyro}) and re-expressing the distribution function as $F_0\left( \boldsymbol{\eta}_{\star}, t \right)$ we obtain, following the steps in Section 4.1, $\partial F_0 / \partial \varphi = 0$ and
\begin{align} \label{kinetic-star}
\dot{\bar{x}}  \frac{\partial F_0}{\partial \bar{x}} + \dot{y}_{\star} \frac{\partial F_0}{\partial y_{\star}}  + \dot{\mu} \frac{\partial F_0}{\partial \mu} + \dot{\varphi} \frac{\partial F_1}{\partial \varphi} =  \mathcal{C}[F_0]   \text{.}
\end{align}
Note that all partial derivatives in (\ref{kinetic-star}) are carried out holding the other variables in $\boldsymbol{\eta}_{\star}$ and $t$ constant. Gyroaveraging (\ref{kinetic-star}) and using the results (\ref{xbardot}), (\ref{ad-invariance}) and (\ref{ystardot}) we obtain the gyrokinetic equation
\begin{align} \label{gyrokinetic-star-collisional}
\left( -\alpha v_{\parallel} \left( U_{\perp}, U, \sigma_{\parallel} \right) - \frac{1}{B} \left\langle \frac{\partial \phi}{\partial y}\left(x_{\varphi}, y_{\star} + v_{\parallel} / \alpha \Omega \right) \right\rangle_{\varphi} \right) \frac{\partial F_0}{\partial \bar{x}} \Bigr\rvert_{y_{\star}, \mu, U} = \left\langle \mathcal{C}[F_0] \right\rangle_{\varphi} \text{,}
\end{align}
where we have also used (\ref{ystar-y-vpar}) to re-express $y$ in terms of $y_{\star}$ and the parallel velocity $v_{\parallel} \left( U_{\perp}, U, \sigma_{\parallel} \right)$. Note that the perpendicular energy $U_{\perp}$ in the argument of $v_{\parallel}$ depends in a complicated way on $\mu$, $y_{\star}$, $\bar{x}$ and $U$. Equation (\ref{gyrokinetic-star-collisional}) only requires the boundary conditions on $\bar{x}$ to solve it. Moreover, it simplifies to 
\begin{align} \label{gyrokinetic-collisionless}
\frac{\partial F_0}{\partial \bar{x}} \Bigr\rvert_{y_{\star}, \mu, U} = 0
\end{align}
when the collisionless limit, corresponding to $\rho_{\text{i}} \ll \alpha \lambda_{\text{mfp}} $, is taken. Equation (\ref{gyrokinetic-collisionless}) is a statement that the distribution function is, to lowest order, only a function of the constants of motion $y_{\star}$, $\mu$ and $U$. This function is determined from the boundary condition at the collisionless magnetic presheath entrance $\bar{x} \rightarrow \infty $,
\begin{align} \label{F0-xbar-infinity-star}
\lim_{\bar{x} \rightarrow \infty} F_0 = F_0^{\infty} \left( y_{\star}, \mu, U \right) \text{ for } \left\langle \dot{\bar{x}} \right\rangle_{\varphi} < 0 \text{,}
\end{align}
and the boundary condition close to the wall, (\ref{bc-F0-xbar-wall}). We therefore have
\begin{align} \label{F-solution}
F_0 = \begin{cases}
F_0^{\infty} \left( y_{\star}, \mu, U \right) & \text{ for \emph{closed} orbits,} \\
0 & \text{ for \emph{open} orbits.}
\end{cases}
\end{align}
This result is similar to the one obtained in reference \cite{Cohen-Ryutov-1998}, but generalized using $y_{\star}$ in order to account for gradients parallel to the wall in a simple way.

\section{Quasineutrality}

In this section, we derive a gyrokinetic quasineutrality equation that is valid in a grazing angle magnetic presheath next to an electron repelling wall. From the equation we obtain, we can, in principle, solve for the electrostatic potential in such a presheath.
In Section 5.1 we write the ion density as an integral of the distribution function expressed in the set of variables $\boldsymbol{\eta}$ or $\boldsymbol{\eta}_{\star}$. This integral is better understood in terms of integrating over the set of curves of the effective potential $\chi$ which allow the ion to be in a closed orbit. Assuming Boltzmann electrons, we obtain the quasineutrality condition in Section 5.2 and simplify it by assuming a collisionless magnetic presheath in Section 5.3. We conclude, in Section 5.4, by making some remarks about the validity of the equations we derived.

\subsection{The ion density integral}

The number density of ions is obtained by integrating the ion distribution function $f$ in velocity space,
\begin{align} \label{ion-density-nongyro}
n_{\text{i}} = \int d^3v f(\vec{r}, \vec{v}) \text{.}
\end{align}
We change variables in the ion density integral (\ref{ion-density-nongyro}) from $\left\lbrace v_x , v_y , v_z \right\rbrace$ to $\left\lbrace \bar{x}, U_{\perp}, U \right\rbrace$ while holding the ion position $\vec{r}$ fixed (the position $\vec{r}$ is determined by $\varphi$, $y$ and $z$ for fixed values of $\bar{x}$, $U_{\perp}$ and $U$). To this end, we calculate the Jacobian of the change of variables. Using (\ref{vy-xbar-def}), (\ref{vx}) and (\ref{vz-U-Uperp}) we obtain the Jacobian $\left| \partial \left( v_x, v_y, v_z \right) / \partial \left( \bar{x}, U_{\perp}, U \right) \right| = \Omega / \left| v_x v_z \right|$, which is, in terms of the new variables, 
\begin{align}
\left| \frac{\partial \left( v_x , v_y , v_z\right)}{\partial \left( \bar{x}, U_{\perp}, U \right)} \right| = \frac{\Omega}{\sqrt{2\left( U_{\perp} - \chi \left( x, \bar{x}, y \right) \right)} \sqrt{2\left( U - U_{\perp} \right)}} \text{.}
\end{align}
The ion density integral therefore becomes, without the integral limits indicated explicitly yet
\begin{align} \label{ion-density-appendix}
n_{\text{i}}\left( \vec{r} \right) = \sum_{\sigma_{\parallel} = \pm 1}   \int d\bar{x} \int \frac{2dU_{\perp}}{ \sqrt{2\left( U_{\perp} - \chi \left( x, \bar{x}, y \right) \right)} }\int \frac{\Omega F_0\left( \bar{x}, y, U_{\perp}, U, \sigma_{\parallel} \right)}{\sqrt{2\left( U - U_{\perp} \right)}} dU \text{,} 
\end{align}
where the summation over the two possible values $\sigma_x = \pm 1$ has simplified to a factor of $2$ because $F_0$ is independent of $\varphi$. If the distribution function were gyrophase dependent, the summation over $\sigma_x$ would be necessary and (\ref{varphi-def}) would be used in order to obtain $\varphi \left(x, U_{\perp}, \bar{x}, \sigma_x \right)$ at each integration point.

The integral limits required in (\ref{ion-density-appendix}) correspond to the phase space boundary between open and closed orbits. In references \cite{Cohen-Ryutov-1998, Holland-Fried-Morales-1993, Gerver-Parker-Theilhaber-1990, Parks-Lippmann-1994}, analytical expressions for the phase space boundary between the closed ion orbits and the open orbits that are quickly lost to the wall are found with different various assumptions. These assumptions are in most cases (except for \cite{Gerver-Parker-Theilhaber-1990}) equivalent to ignoring or simplifying the effect of strong orbit distortion.
In what follows, we identify the phase space boundary between open and closed orbits in a way that is general (allowing strong orbit distortion) and allows its numerical determination. The procedure is similar to the one presented by Gerver, Parker and Theilhaber in reference \cite{Gerver-Parker-Theilhaber-1990}, but includes the velocity dimension parallel to the magnetic field, weak gradients parallel to the wall, and allows for the presence of an effective potential maximum near the wall. Such a maximum was never present in \cite{Gerver-Parker-Theilhaber-1990} because the authors were studying the different problem of a plasma with a magnetic field parallel to the wall. Hence, the wall was charged positively (see discussion in Section 2 on our choice $\alpha \gg \sqrt{m_{\text{e}} / m_{\text{i}}}$) and was ion repelling. 

The integral in (\ref{ion-density-appendix}) is carried out at a fixed position $\vec{r}$. In this section, we choose to denote the effective potential $\chi$ as a function of the variable $s$ in order to distinguish the position $x$ at which we are evaluating the ion density (and hence carrying out the integral in (\ref{ion-density-appendix})) from the positions $s$ that the particle occupies in its lowest order orbit. Obviously $x$ is just one of the many possible values that $s$ can take. The limits for $\bar{x}$ are dictated by the restriction $s \geq 0$ that the presence of a wall at $s = 0$ implies on the effective potential. For some values of $\bar{x}$, the effective potential does not allow a closed orbit that crosses the position $\vec{r}$. The orbits with these values of $\bar{x}$ are \emph{open} at position $\vec{r}$, and by the boundary condition (\ref{bc-F0-xbar-wall}) the distribution function for such orbits is zero.

\begin{figure}
\centering
\includegraphics[scale=0.3]{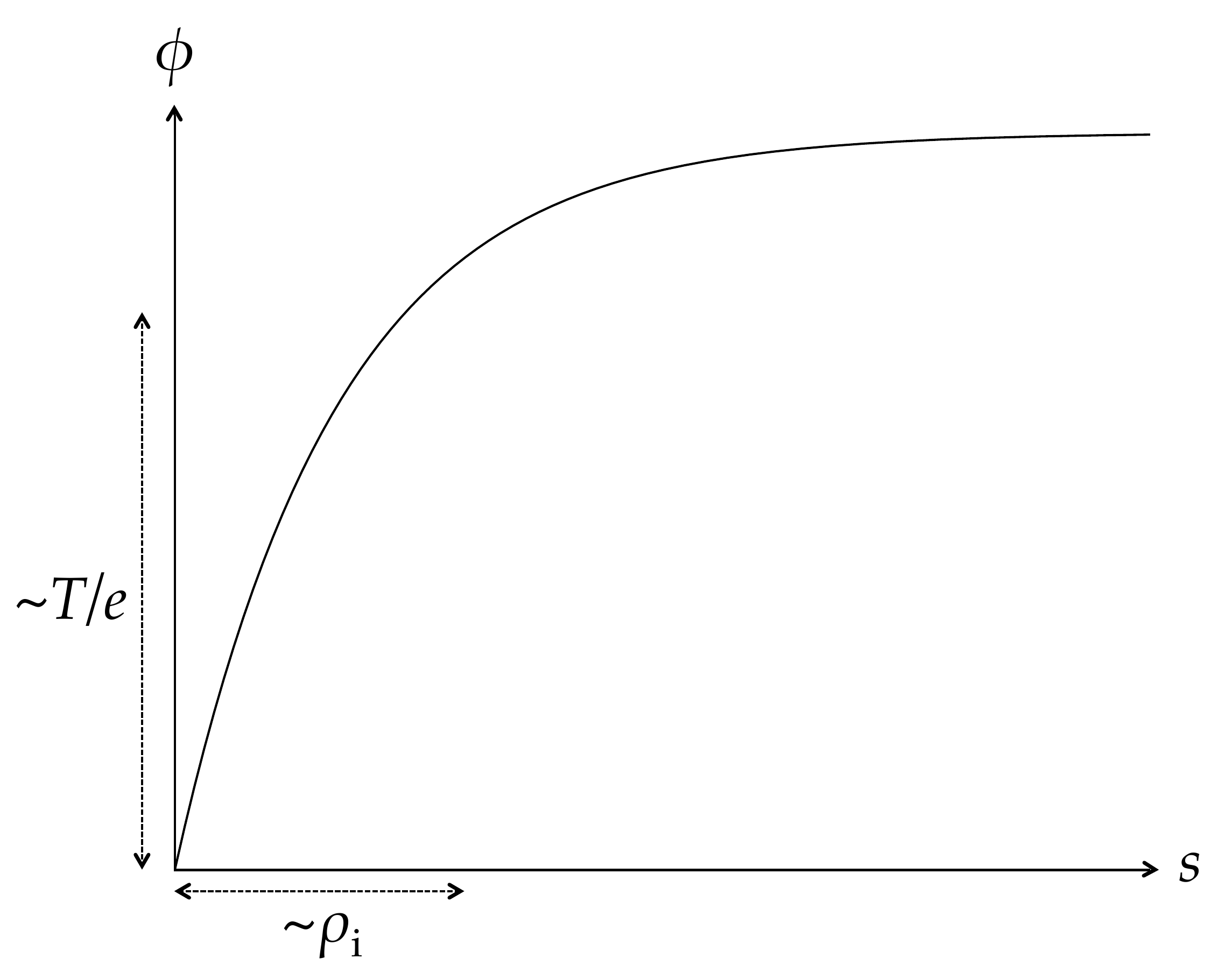}
\caption{The shape of the electrostatic potential that we expect in the magnetic presheath. The electric field in the magnetic presheath (which is the gradient of the electrostatic potential) increases as the wall is approached.}
\label{figure-potential-shape-presheath}
\end{figure}

Figure \ref{figure-potential-shape-presheath} shows the expected shape of the presheath electrostatic potential. Electric fields outside the magnetic presheath are weak in our ordering, so that $\partial \phi / \partial s \left( s \rightarrow \infty , y\right) \simeq 0$. The effective potential $\chi$ must therefore be unbounded at infinity for finite $\bar{x}$, 
\begin{align} \label{eff-pot-bound}
\frac{\partial \chi}{\partial s} \left( s \rightarrow \infty , \bar{x}, y \right) \simeq \Omega^2 \left( s - \bar{x} \right) > 0 \text{,}
\end{align}
leading always to a bounce point for sufficiently large $x$.
With the knowledge that the effective potential is unbounded at infinity, in order to have a closed orbit crossing the position $\vec{r}$ at which we are calculating the integral, the effective potential must be larger than its value at $\vec{r}$ for some value of $s$ between the particle position $s = x$ and the wall at $s = 0$,
\begin{align}
\chi \left(s, y, \bar{x} \right) > \chi \left( x, y,  \bar{x}\right) \text{ for some or all } s \in \left[0, x\right) \text{.}
\end{align} 
Explicitly, this is (after dividing through by $\Omega^2$)
\begin{align}
\frac{1}{2} \left( s - \bar{x} \right)^2 + \frac{\phi \left( s, y \right)}{\Omega B} > \frac{1}{2}  \left( x - \bar{x} \right)^2 + \frac{\phi \left( x, y \right)}{\Omega B} \text{,}
\end{align}
which reduces to
\begin{align}
\bar{x} \left( x - s \right) > \frac{1}{2} \left( x^2 - s^2 \right) + \frac{\phi \left( x, y \right) - \phi \left( s, y \right)}{\Omega B} \text{.}
\end{align}
This leads to the closed orbit condition
\begin{align} \label{condition-closed}
\bar{x} > g\left( s, x, y \right) \equiv  \frac{1}{2}\left( x + s \right) + \frac{\phi \left( x, y \right) - \phi \left( s, y \right)}{\Omega B \left( x - s \right)} \text{ for some or all } s \in \left[0, x\right) \text{.}
\end{align}

The inequality in (\ref{condition-closed}) is the condition that must be satisfied by the orbit position $\bar{x}$ in order for there to be some closed orbits with such value of $\bar{x}$ that cross the point $x$ (at which we evaluate the integral (\ref{ion-density-appendix})). The minimum value of $\bar{x}$, $\bar{x}_{\text{m}} \left( x, y \right)$, that satisifies this condition is therefore given by minimizing the function $g\left( s, x, y \right)$ over the interval $\left[0, x\right)$,
\begin{align}
\bar{x}_{\text{m}} \left( x, y \right) = \min_{s \in \left[0, x \right)} g\left(s, x, y\right) \text{.}
\end{align}
Note that, from (\ref{condition-closed}), $g\left( s, x, y\right) > 0$ because $s < x$ and the electrostatic potential $\phi$ is increasing with the distance to the (negatively charged) wall (see Figure \ref{figure-potential-shape-presheath}). This implies that $\bar{x}_{\text{m}} \left( x , y \right) > 0$. No closed orbit can exist for $\bar{x} < \bar{x}_{\text{m}} \left( x, y \right)$, which leads to the integration domain for $\bar{x}$ in (\ref{ion-density-appendix}) being $\left[ \bar{x}_{\text{m}} \left(x, y \right), \infty \right] $.

\begin{figure}
\centering
\includegraphics[width=0.45\textwidth]{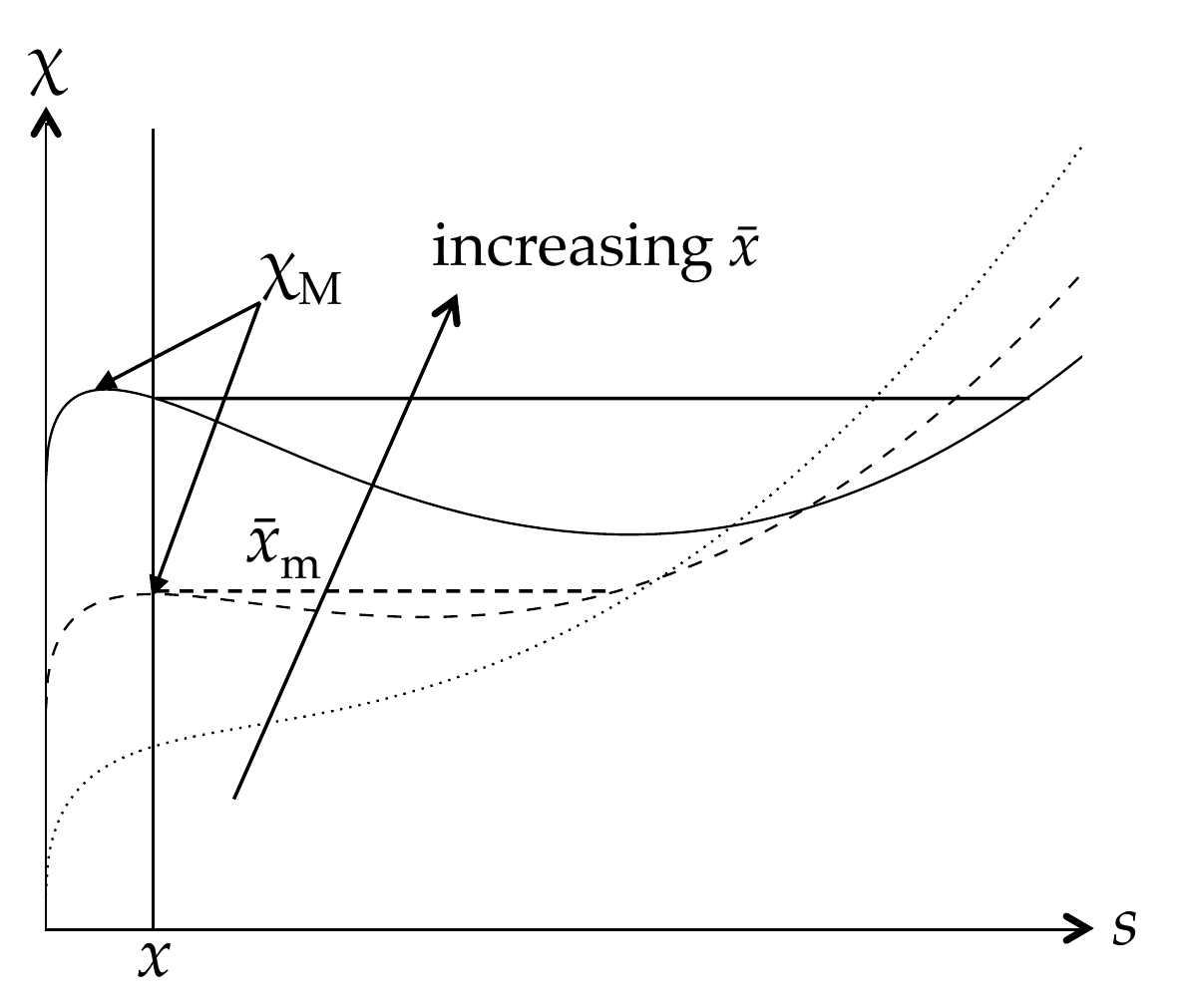}
\includegraphics[width=0.45\textwidth]{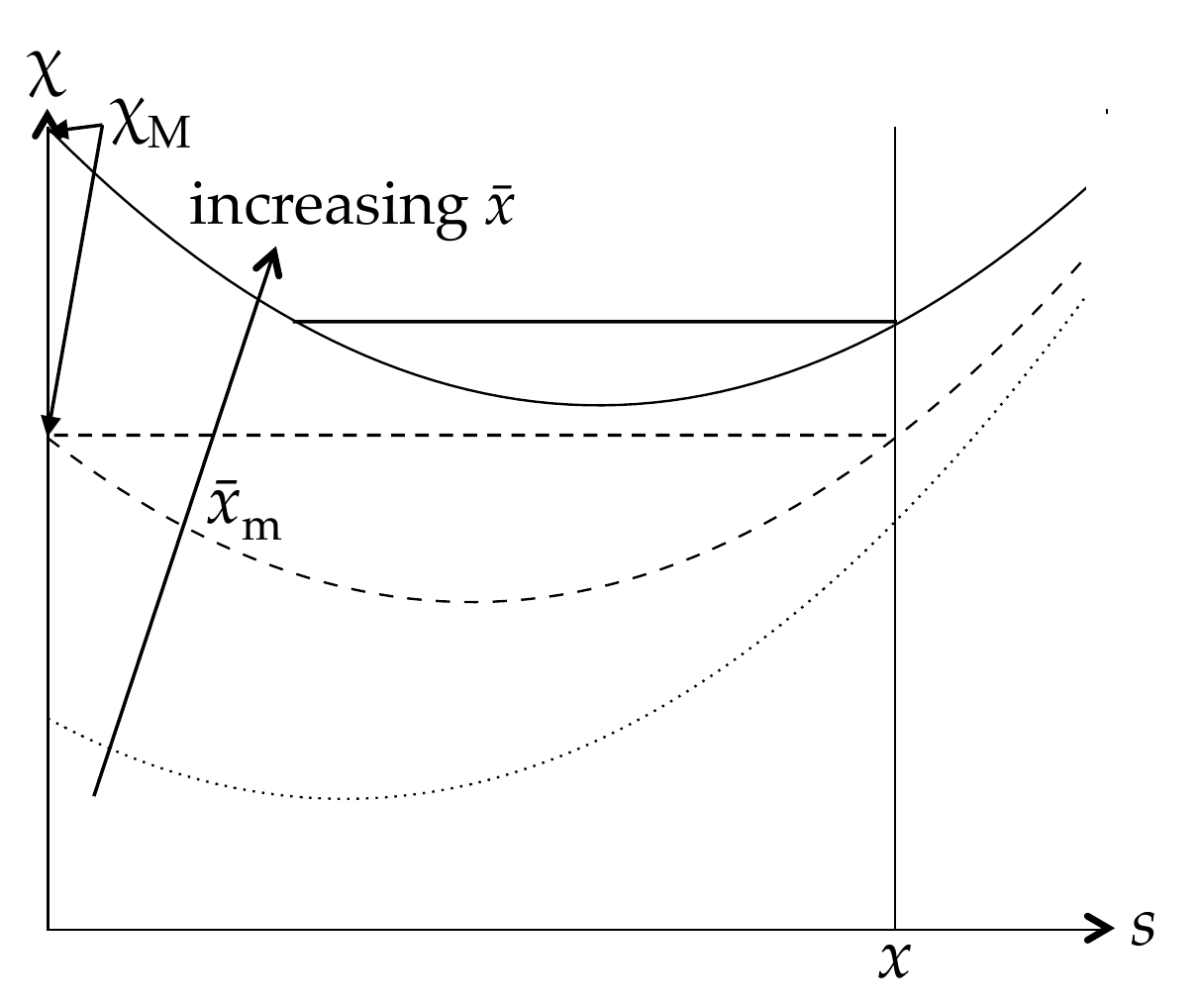} \newline
 (a) \hspace{6cm} (b)
\caption{Two sets of effective potential curves, each set with three different curves $\chi \left(s, \bar{x}, y \right)$ plotted as a function of $s$ that correspond to a different value of the orbit position $\bar{x}$. In (a), effective potential curves with a local maximum near the wall, which arises if the electric field is sufficiently strong there, are shown. In (b), effective potential curves without a maximum are shown. The solid curves are associated to orbits with position $\bar{x} > \bar{x}_{\text{m}} \left( x, y \right)$. Dashed curves correspond to orbit position $\bar{x}_{\text{m}} \left( x, y \right)$ which is associated with the presence of only one semi-closed orbit 
that passes through $x$, while dotted ones correspond to orbits at position $\bar{x} < \bar{x}_{\text{m}} \left( x, y \right)$, which are all open if they are to cross point $x$. The horizontal lines are associated with the minimum perpendicular energy required for a closed orbit to lie at position $\bar{x}$ and cross the point $x$, equal to $\chi \left( x, \bar{x}, y \right)$. The maximum value of $\chi$ between the point $s = x$ and the wall $s = 0$ is $\chi_{\text{M}} \left( \bar{x}, y \right)$, marked for the closed (and semi-closed) orbit curves. In (a) $\chi_{\text{M}} \left(\bar{x}, y \right)$ corresponds to the local effective potential maximum between $s=x$ and $s=0$, while in (b) $\chi_{\text{M}} = \chi\left(0, \bar{x}, y \right)$.}
\label{figure-xbarmin-condition}
\end{figure}

For a given $x$ and $y$, we define the largest value of the effective potential between the position $x$ and the wall as
\begin{align}
\chi_{\text{M}} \left( \bar{x}, y \right) = \max_{s \in \left[ 0, x \right)} \chi \left(s, \bar{x}, y \right) \text{.}
\end{align}
This is illustrated in Figure \ref{figure-xbarmin-condition}. For an orbit with $\bar{x} > \bar{x}_{\text{m}} \left( x, y \right)$, we are assured that $\chi_{\text{M}} \left( \bar{x}, y \right) > \chi\left( x, \bar{x}, y \right)$ which leads to the presence of closed orbits with that orbit position. However, any orbit which has perpendicular energy $U_{\perp} > \chi_{\text{M}} \left( \bar{x}, y \right)$ is open, with its trajectory intersecting the wall. Hence, such an orbit does not contribute to the ion density to lowest order and is not integrated over in (\ref{ion-density-appendix}). Values $U_{\perp} < \chi\left( x, \bar{x}, y \right)$ are not allowed because the particle must have enough perpendicular energy to find itself at the point $x$ with orbit position $\bar{x}$. This leads to the integration domain of $U_{\perp}$ in (\ref{ion-density-appendix}) being $U_{\perp} \in  \left[\chi \left( x, \bar{x}, y \right), \chi_{\text{M}} \left(\bar{x}, y \right) \right]$. The domain of integration of $U$ is only constrained by $U \geq U_{\perp}$, leading to the domain $U \in \left[ U_{\perp}, \infty \right] $. 

The ion density integral in (\ref{ion-density-appendix}) can be re-written with the correct limits explictly indicated
\begin{align} \label{ion-density}
n_{\text{i}} = \sum_{\sigma_{\parallel} = \pm 1} \int_{\bar{x}_m \left( x , y \right) }^{\infty} d\bar{x} \int_{\chi\left( x, \bar{x}, y \right)}^{\chi_{\text{M}} \left( \bar{x}, y \right)}  \frac{2\Omega dU_{\perp}}{\sqrt{2\left( U_{\perp} - \chi \left( x, \bar{x} \right) \right)}} \int_{U_{\perp}}^{\infty} \frac{F_0 \left(\bar{x}, y, U_{\perp} , U, \sigma_{\parallel} \right)}{\sqrt{2\left( U - U_{\perp}\right)}} dU \text{.}
\end{align}
The limits ensure that we only integrate over the set of closed orbits, in accordance with the boundary condition at the wall (\ref{bc-F0-xbar-wall}).

\subsection{Quasineutrality equation in general form}

The magnetic presheath is quasineutral ($\lambda_D \ll \rho_{\text{i}}$): $Z n_{\text{i}} = n_{\text{e}}$. Because we assume an electron repelling wall, $\alpha \gg \sqrt{m_{\text{e}}/m_{\text{i}}}$, the electrons are in thermodynamic equilibrium, and are therefore Boltzmann distributed throughout the magnetic presheath, \begin{align} \label{electrons-adiabatic}
n_{\text{e}} = n_{\text{e}\infty} \left( y \right) \exp{\left(\frac{e\left( \phi \left(x, y \right) - \phi_{\infty} \left( y \right) \right)}{T_e }\right)} \text{.}
\end{align}
In (\ref{electrons-adiabatic}), $n_{\text{e}\infty} \left(y \right)$ and $\phi_{\infty} \left( y \right)$ are the electron density and the electrostatic potential at the magnetic presheath entrance ($x \rightarrow \infty$). While it is true that the truncation of the Maxwellian electron distribution function, due to the high energy electrons reaching the wall instead of being reflected, leads to a correction to the Boltzmann distribution (\ref{electrons-adiabatic}) \cite{Ingold-1972}, this correction is of order $n_{\text{e},\infty}\sqrt{m_{\text{e}}/m_{\text{i}}}$. Our equations are lowest order in $\sqrt{m_{\text{e}}/m_{\text{i}}}$, therefore we use equation (\ref{electrons-adiabatic}), instead of any corrected version of it, to calculate the electron density.

With the electron density of (\ref{electrons-adiabatic}) the quasineutrality equation becomes, to lowest order,
\begin{align} \label{quasineutrality-general}
& n_{\text{e}\infty} \left(y \right) \exp{\left(\frac{e\left( \phi \left( x, y \right) - \phi_{\infty} \left(y \right) \right)}{T_e }\right)} = \nonumber \\
& Z \sum_{\sigma_{\parallel} = \pm 1} \int_{\bar{x}_m \left( x, y \right) }^{\infty} d\bar{x} \int_{\chi\left( x, \bar{x}, y \right)}^{\chi_{\text{M}} \left(\bar{x}, y \right)}  \frac{2\Omega dU_{\perp}}{\sqrt{2\left( U_{\perp} - \chi \left( x, \bar{x} \right) \right)}} \int_{U_{\perp}}^{\infty} \frac{F_0(\bar{x}, y, U_{\perp} , U, \sigma_{\parallel})}{\sqrt{2\left( U - U_{\perp}\right)}} dU \text{,}
\end{align}
where $F_0$ is the solution to the gyrokinetic equation (\ref{gyrokinetic-full-explicit}) throughout the magnetic presheath. The form of the electrostatic potential $\phi$ allows to solve for $F_0$ and dictates what the values of $\bar{x}_m \left( x, y \right) $ and $\chi_{\text{M}} \left( \bar{x}, y \right)$ are. The electrostatic potential must be such that the quasineutrality equation (\ref{quasineutrality-general}) is satisfied to lowest order.

If we express the distribution function using the set of variables $\boldsymbol{\eta}_{\star}$, the quasineutrality equation takes the same form, except that $F_0$ is a different function $F_0 \left( \bar{x}, y_{\star}, \mu , U, \sigma_{\parallel} \right)$ obtained by solving the gyrokinetic equation (\ref{gyrokinetic-star-collisional}). The coordinates $\mu$ and $y_{\star}$ must be obtained at each integration point using equations (\ref{mu(Uperp,xbar)}) (updated to include $y$ dependence, see (\ref{mu(Uperp,xbar,y)-appendix})) and (\ref{ystar-y-vpar}).

\subsection{Quasineutrality equation simplified by taking $\rho_{\text{i}} \ll \alpha \lambda_{\text{mfp}}$}

In the weak collisionality limit ($\rho_{\text{i}} \ll \alpha \lambda_{\text{mfp}}$), the distribution function expressed in terms of the set of variables $\boldsymbol{\eta}_{\star}$ is given by (\ref{F-solution}). Quasineutrality (\ref{quasineutrality-general}) therefore simplifies to
\begin{align} \label{quasineutrality-general-star}
& n_{\text{e}\infty} \left( y \right) \exp{\left(\frac{e\left( \phi \left( x, y \right) - \phi_{\infty} \left( y \right) \right)}{T_e }\right)} = \nonumber \\
& Z \sum_{\sigma_{\parallel} = \pm 1} \int_{\bar{x}_m \left(x, y \right)}^{\infty} d\bar{x}  \int_{\chi\left( x, \bar{x}, y \right)}^{\chi_{\text{M}} \left( \bar{x}, y \right)} \frac{2\Omega dU_{\perp}}{\sqrt{2\left( U_{\perp} - \chi \left( x, \bar{x} \right) \right)}} \int_{U_{\perp}}^{\infty} \frac{F_0^{\infty}\left(y_{\star}, \mu , U\right))}{\sqrt{2\left( U - U_{\perp}\right)}}  dU \text{.}
\end{align}
This equation may, in principle, be used to solve for the electrostatic potential $\phi \left( x , y \right)$. This amounts to finding a potential for which (\ref{quasineutrality-general-star}) is satisfied.% at each integration point.

In a similar way to what was proposed by Cohen and Ryutov \cite{Cohen-Ryutov-1998}, the collisionless magnetic presheath can be solved, using the quasineutrality condition above, by means of an iterative procedure. Here, we make explicit use of the new variables in the quasineutrality equation and account for $y$ dependence in a natural way. In order to solve for the self-consistent electrostatic potential in the magnetic presheath, the electrostatic potential profile in the $y$ direction must be known, as a boundary condition, at the entrance of the magnetic presheath, $x \rightarrow \infty$. This is $\phi_{\infty} \left( y \right) = \phi \left( x \rightarrow \infty, y \right)$. For some guessed potential $\phi \left( x, y \right)$, the integral on the RHS of (\ref{quasineutrality-general-star}) can be computed numerically. If it is possible to find a good enough initial guess of $\phi \left( x, y \right)$ such that the difference between the RHS and the LHS of (\ref{quasineutrality-general-star}) is small, one can correct $\phi \left(x, y \right)$ to make such a difference even smaller, and this procedure can be iterated until convergence.

Such an iteration procedure could be used to solve the simpler problem with negligible parallel gradients ($\delta \ll \alpha$). In such a problem, the distribution function is expected to have a negligible dependence on $y$, and therefore on $y_{\star}$. We can therefore drop the $y_{\star}$ and $y$ dependences from (\ref{quasineutrality-general-star}), so the resulting quasineutrality equation is
\begin{align} \label{quasineutrality-simple}
& n_{\text{e}\infty} \exp{\left(\frac{e\left( \phi \left( x \right) - \phi_{\infty} \right)}{T_e }\right)} = \nonumber \\
& Z  \int_{\bar{x}_m \left(x\right)}^{\infty} d\bar{x}  \int_{\chi\left( x, \bar{x} \right)}^{\chi_{\text{M}} \left( \bar{x} \right)} \frac{2\Omega dU_{\perp}}{\sqrt{2\left( U_{\perp} - \chi \left( x, \bar{x} \right) \right)}} \int_{U_{\perp}}^{\infty} dU \frac{F_0^{\infty}\left(\mu , U\right)}{\sqrt{2\left( U - U_{\perp}\right)}} \text{.} 
\end{align}
We have removed the summation over $\sigma_{\parallel}$ because $\sigma_{\parallel} = - 1 $ (particles moving away from the wall) is not allowed at $\bar{x} \rightarrow \infty$, and ions do not bounce back (they are slowly being accelerated towards the wall by the small component of the electric field which is parallel to the magnetic field, see (\ref{vparalleldot})). In (\ref{quasineutrality-simple}), $n_{\text{e}\infty}$ and $\phi_{\infty}$ are constant.

\subsection{Remarks about the validity of our equations}
To conclude this section, we make some important remarks regarding the validity of the equations presented in this paper. 
\subsubsection{The gyrokinetic approximation}
~\newline
The way we proceeded to construct gyrokinetics assumed that the gyrofrequency, modified by the strongly varying electric field, is much larger than the typical timescale of the particle motion. We expect this ordering to hold for closed orbits with $\chi_{\text{M}}\left(\bar{x}, y \right) - U_{\perp} \gg \alpha v_{\text{t,i}}^2 $. For $\chi_{\text{M}}\left(\bar{x}, y \right) - U_{\perp} \lesssim \alpha v_{\text{t,i}}^2 $, particles leave the effective potential well in a time of order $1/ \Omega$ which is comparable to the typical gyroperiod, therefore we cannot separate the timescale of the problem from the orbital timescale. We are justified in including the ion orbits up to $U_{\perp} = \chi_{\text{M}} \left( \bar{x}, y \right)$ in the closed orbit density (\ref{ion-density}) because this is accurate to lowest order in $\alpha$. The correction to the density due to the orbits with $\chi_{\text{M}}\left(\bar{x}, y \right) - U_{\perp} \lesssim \alpha v_{\text{t,i}}^2 $ is smaller. 

\subsubsection{Potential drop across the magnetic presheath} ~\newline
It turns out that ignoring completely the contribution of the open orbits to the total ion density is problematic if we wish to determine the exact potential drop across the magnetic presheath, or the behaviour of the potential close to the wall. The problem lies in the fact that, to every order in $\alpha$, the density of ions in a closed orbit is zero at $x=0$. This can be seen by noting that an ion at $x=0$ has either crossed the effective potential maximum (and is therefore open) or has $ U_{\perp} = \chi_{\text{M}} \left( \bar{x}, y \right)$ (such as in Figure \ref{figure-xbarmin-condition}(b), dashed line and curve), and therefore the upper and lower limits of the integral over perpendicular energy in the ion density (\ref{ion-density}) are the same. Ignoring the $y$ dependence for simplicity and solving the quasineutrality equation (\ref{quasineutrality-simple}) at $x=0$ results in $n_{\text{e}\infty} \exp\left( e \phi_{\text{MPS}} / T_{\text{e}} \right) = 0$, with $\phi_{\text{MPS}} = \phi \left( 0 \right) - \phi_{\infty}$. Therefore, an infinite potential drop is obtained across the magnetic presheath, $\phi_{\text{MPS}}  \rightarrow - \infty $. This is not true in practice, as the total potential drop across the sheath-presheath system is certainly finite \cite{Stangeby-book}. 

If we could quantify the open orbit density, and keep it in (\ref{quasineutrality-simple}) when solving near the wall, we would obtain the true potential drop and the correct variation of the potential near $x=0$. The density of ions in open orbits is small, of size $\sim \alpha n_{\text{e}\infty}$ or $ \sim \sqrt{\alpha} n_{\text{e}\infty}$ depending on whether there is or is not, respectively, an effective potential maximum in front of the wall for most ion orbits (see Figure \ref{figure-xbarmin-condition}) \cite{Cohen-Ryutov-1998}.
Using the scaling for the open orbit ion density $n_{\text{i,open}} \sim \alpha^{p} n_{\text{e}\infty}$ with $p = 1/2$ or $p=1$, we can derive the scaling for the expected potential drop in the magnetic presheath. By using quasineutrality (\ref{quasineutrality-simple}) with the open orbit density at $x=0$, we have 
\begin{align}
n_{\text{e}\infty} \exp \left( \frac{ e \phi_{\text{MPS}} }{ T_{\text{e}}} \right) \sim \alpha^p n_{\text{e}\infty} \text{,}
\end{align}
which immediately implies that the potential drop across the magnetic presheath is
\begin{align} \label{eq:MPS-pot-drop}
\phi_{\text{MPS}} = \phi\left( 0 \right) - \phi_{\infty} \sim \frac{T}{e} p \ln \alpha \text{.}
\end{align}
The scaling (\ref{eq:MPS-pot-drop}) is equivalent to the one that is derived when using fluid equations \cite{Chodura-1982, Riemann-1994, Loizu-2012}. 

A future publication will include an analysis of the transition from closed to open ion orbits in the region of the magnetic presheath right next to the wall, and provide an expression for the open orbit density. Such an expression can then be used to calculate $\phi_{\text{MPS}}$.

\subsubsection{Multiple ion species}
~\newline
Throughout this work we have assumed a single ion species. However, the gyrokinetic equations we derived are valid separately for each ion species in a multi-species system, and the quasineutrality equation can be generalized to include more than one ion species (by just adding the density integral of the additional species). Generalizing to a system with more than one ion species may be useful to account for the presence of Deuterium and Tritium isotopes in roughly equal amounts near the divertor targets of potential future fusion devices.

\section{Conclusion}

We have developed a gyrokinetic treatment for ions in a grazing angle magnetic presheath such as the one present next to a tokamak divertor target.
In our treatment, we made use of the smallness of the angle $\alpha$ between the magnetic field and the wall, and the small ratio $\delta$ of ion Larmor radius to the characteristic length scale of variations parallel to the wall. Such variations are assumed to be set by the cross-field width of turbulent structures reaching the divertor target from the SOL. The ordering $\delta \sim \alpha$ allows us to include in our equations gradients parallel to the wall with scale length $l = \rho_{\text{i}} / \delta \gg \rho_{\text{i}}$. This enables our equations to describe realistic divertor target magnetic presheaths where the SOL width and the presence of turbulent structures, such as SOL filaments impinging on the divertor target, imply the presence of gradients parallel to the wall. 
Orbit distortion due to the spatially changing electric field normal to the wall was retained to lowest order in our treatment, while the magnetic field was assumed to be constant in space and time.% due to the very small plasma $\beta$ parameter.

Within these approximations, we showed that the distribution function is independent of gyrophase to lowest order, and we derived the gyrokinetic equations (\ref{gyrokinetic-full-explicit}) and (\ref{gyrokinetic-star-collisional}) (using the two different sets of variables $\boldsymbol{\eta}$ and $\boldsymbol{\eta}_{\star}$ introduced in equations (\ref{xi-eta0}) and (\ref{eta-etastar})) correct to first order in $\alpha \sim \delta$. In (\ref{gyrokinetic-full-explicit}) and (\ref{gyrokinetic-star-collisional}) we retained a collision operator whose form we did not attempt to discuss, but which allows the magnetic presheath to be collisional, $\alpha \lambda_{\text{mfp}} \sim \rho_{\text{i}}$. We also derived the quasineutrality equation (\ref{quasineutrality-general}), which assumes Boltzmann electrons. The gyrokinetic and quasineutrality equations can be used to solve self-consistently for the ion distribution function and the electrostatic potential in the magnetic presheath, ignoring the contribution to the density of open ion orbits hitting the wall. Quantifying and treating the effect of open orbits will be the subject of a future publication. 

If the limit of a collisionless magnetic presheath ($\rho_{\text{i}} \ll \alpha \lambda_{\text{mfp}}$) is taken, we obtain a simple form for the gyrokinetic equation which implies that the distribution function in terms of the set of variables $\boldsymbol{\eta}_{\star}$ is constant throughout the magnetic presheath, and is therefore equal to the distribution function at the presheath entrance, (\ref{F-solution}). This solution arises because the set of variables $\boldsymbol{\eta}_{\star}$ includes three quantities, which are the total energy $U$, the adiabatic invariant $\mu$ and the coordinate $y_{\star}$ (proportional to the $z$ component of the canonical momentum), which are conserved to lowest order in $\alpha \sim \delta$ over the timescale of the problem, the magnetic presheath timescale $t_{\text{MPS}}$. Using the quasineutrality condition (\ref{quasineutrality-general-star}), we outlined how an iteration procedure could solve for the electrostatic potential in a collisionless magnetic presheath, taking into account the turbulent electric fields in a natural way by exploiting the coordinate $y_{\star}$.

The assumption of a collisionless magnetic presheath provides a good starting point for the understanding of the plasma-wall boundary from a kinetic point of view, as well as being justified by the estimates we have made for the length scales of the plasma-wall boundary layers in a typical tokamak plasma. Once the collisionless magnetic presheath (with scale length $x \sim \rho_{\text{i}}$) is solved for self-consistently and rigorously, for some boundary conditions at its entrance, it is natural to ask the question of what the correct kinetic boundary conditions at its entrance are, and whether the solution for the distribution function has the correct behaviour at the magnetic presheath exit, which has to be matched to the Debye sheath entrance. To answer the first of these questions, a solution of the ion kinetic equation in the collisional region, $x \sim \alpha \lambda_{\text{mfp}}$ (the layer furthest from the wall in Figure \ref{figure-boundary-layers}) is required, which can then be matched to the magnetic presheath.
To address the second question, the solution at the magnetic presheath exit must satisfy the kinetic conditions at the Debye sheath entrance (provided that such a Debye sheath is present \cite{Coulette-Manfredi-2016, Stangeby-2012}) first derived by Harrison and Thompson in 1959 \cite{Harrison-Thompson-1959} and discussed in detail in the review by Riemann \cite{Riemann-review}. 

It is worth noting that we are working in the context of an electron repelling divertor target, and hence adiabatic (or Boltzmann) electrons. This assumption is not valid at sufficiently small angles $\alpha \sim 1^{\circ}$, as pointed out in Section 2. For current tokamak machines, it is, however, a reasonable assumption. If the magnetic field to wall angle were made sufficiently small in future machines, a collisional model accounting for fully kinetic electrons (similar to the one in reference \cite{Gerver-Parker-Theilhaber-1990} for $\alpha = 0$) would be necessary to study the physics of the plasma-wall boundary at divertor targets correctly.

\ack{A. G. would like to thank J. Ball for discussions on the geometry and the physics of tokamaks, and G. Hammett for correcting an inaccuracy on one of the boundary conditions to the gyrokinetic equation. This work has been carried out within the framework of the EUROfusion Consortium and has received funding from the Euratom research and training programme 2014-2018 under grant agreement No 633053 and from the RCUK Energy Programme [grant number EP/I501045]. The views and opinions expressed herein do not necessarily reflect those of the European Commission.}

\appendix

\section{Estimates for the typical width of the different plasma-wall boundary layers in a tokamak} \label{appendix:widths}

In order to estimate the typical width of the different plasma-wall boundary layers (see Figure \ref{figure-boundary-layers}), we extract data from reference \cite{Militello-Fundamenski-2011}, which contains a comparison of some important parameters between different tokamaks. We use JET data because it is the most relevant for fusion. From Tables 1 to 5 of \cite{Militello-Fundamenski-2011}, we estimate the magnetic field,
\begin{align}
B \sim 2 \text{ T,}
\end{align}
the electron and ion temperatures (by taking $T_{\text{e}} \sim T_{\text{i}} \sim T$),
%take an average of the plasma density and electron temperature in the near and far SOL of JET, for both H and L mode, to get
\begin{align}
T \sim 50 \text{ eV} = 8 \times 10^{-18} \text{ J} \text{,}
\end{align}
and the electron density,
\begin{align} \label{estimate-density}
n_{\text{e}} \sim 1 \times 10^{19} \text{ m}^{-3} \text{.}
\end{align}
For a deuterium plasma, the ion mass is $m_{\text{i}} \sim 3 \times 10^{-27} \text{ kg}$, which leads to the estimate
\begin{align}
v_{\text{t,i}} = \sqrt{\frac{2T_{\text{i}}}{m_{\text{i}}}} \sim 7 \times 10^{4} \text{ ms}^{-1}
\end{align}
for the ion thermal velocity.
The ion gyrofrequency is estimated as
\begin{align}
\Omega = \frac{eB}{m_{\text{i}}} \sim 1 \times 10^8 \text{ s}^{-1}  \text{,}
\end{align}
which leads to the estimate for the ion gyroradius
\begin{align}
\rho_{\text{i}} \sim \frac{v_{\text{t,i}}}{\Omega} \sim 7 \times 10^{-4} \text{ m} = 0.7 \text{ mm} \text{.}
\end{align}
The Debye length is (with $\epsilon_0 = 8.85 \times 10^{-12} \text{ Fm}^{-1}$ the vacuum permittivity)
\begin{align}
\lambda_{\text{D}} = \sqrt{\frac{\epsilon_0 T_{\text{e}}}{n_{\text{e}}e^2}} \sim  2 \times 10^{-5} \text{ m} = 0.02 \text{ mm} \text{.}
\end{align}

The ion mean free path is estimated for the two dominant collision processes occurring close to the divertor target: Coulomb collisions and charge exchange. For Coulomb collisions, we determine the frequency of ion-ion collisions as (from the NRL Plasma Formulary \cite{NRL-plasma-formulary})
\begin{align} \label{estimate-Coulomb-collisions}
\nu_{\text{ii}} \sim \frac{4\sqrt{\pi}}{3} \frac{e^4n_{\text{i}} \ln \Lambda}{\left(4\pi \epsilon_0 \right)^2 m_{\text{i}}^{1/2} T_{\text{i}}^{3/2} } \sim  2 \times 10^{4} \text{ s}^{-1}\text{,} 
\end{align}
where $\ln \Lambda \sim 15$ is the Coulomb logarithm for ion-ion collisions.
Therefore, the collisional mean free path is
\begin{align} \label{estimate-meanfreepath}
\lambda_{\text{mfp,ii}} \sim \frac{v_{\text{t,i}}}{\nu_{\text{ii}}} \sim  4 \text{ m} = 4 \text{ m.}
\end{align}
We evaluate $\alpha \lambda_{\text{mfp,ii}}$ for $\alpha \sim 0.1$ to obtain
\begin{align}
\alpha \lambda_{\text{mfp,ii}} \sim 0.4 \text{ m.}
\end{align}
For charge exchange we use the value of $ 5 \times 10^{-14} \text{ m}^3\text{s}^{-1}$ (extracted from reference \cite{Havlickova-Fundamenski-2013}) as an approximate rate coefficient for the reaction at $T_{\text{i}} \sim 50 \text{ eV}$. We then multiply this by an estimate of the neutral density, $n_{\text{n}} \sim 10^{18} \text{ m}^{-3}$,
in order to obtain the charge exchange collision frequency
\begin{align}
\nu_{\text{cx}} \sim 5 \times 10^4 \text{ s}^{-1}\text{.}
\end{align}
From this collision frequency, we obtain a mean free path that is slightly smaller than the Coulomb collision one,
\begin{align} \label{estimate-meanfreepath-2}
\lambda_{\text{mfp,cx}} \sim 1 \text{ m} = 1 \text{ m.}
\end{align}
The estimate for the width of the collisional layer becomes
\begin{align}
\alpha \lambda_{\text{mfp,cx}} \sim 0.1 \text{ m.}
\end{align}

A few comments on these numbers and the scalings associated with them are worth making. The scaling $\rho_{\text{i}} / \lambda_{\text{D}} \sim 40 \gg 1$ which implies a quasineutral magnetic presheath is, in our opinion, robust. The dependence on density and temperature of the ion gyroradius and Debye length is weak, with $\rho_{\text{i}} \propto \sqrt{T_{\text{i}}}$ and $\lambda_{\text{D}} \propto \sqrt{T_{\text{e}}/n_{\text{e}}}$, so the error associated with both these estimates is small. If one of $T_{\text{i}}$, $T_{\text{e}}$ or $n_{\text{e}}$ is wrong by a factor of $10$, the corresponding estimate for $\rho_{\text{i}}$ or $\lambda_{\text{D}}$ will be wrong by a factor of $\sim 3$ only. 

The discussion about the length of the collisional layer is more complex. We have assumed Coulomb or charge exchange collisions in order to make our estimates (\ref{estimate-meanfreepath}) and (\ref{estimate-meanfreepath-2}), which resulted in $\alpha \lambda_{\text{mfp}} / \rho_{\text{i}} \sim 100 \gg 1$ using the slightly more conservative charge exchange estimate. This seems to  favour a collisionless model for the magnetic presheath and a scale separation between the magnetic and collisional layers. However, we note that $\lambda_{\text{mfp,ii}} \propto T_{\text{i}}^2$, so that if the ion temperature were smaller than the estimated temperature by a factor of $10$, this separation of scales would no longer be valid due to Coulomb collisions becoming important in the magnetic presheath. Moreover, the charge exchange frequency depends linearly on the neutral density close to divertor targets, which we estimated crudely. Such remarks warrant care towards the idea of a completely collisionless magnetic presheath and the separation into the three layers of Figure \ref{figure-boundary-layers}, even though it is still a physically motivated and theoretically attractive way to model the plasma-wall boundary.

\section{Orderings for the length scales parallel to the wall and the turbulent timescales} \label{appendix:turbulence}

In this appendix, we discuss the orderings for the length scales parallel to the wall in the magnetic presheath, and the characteristic timescale for changes to occur due to turbulence. In Appendix B.1, we recover the orderings (\ref{xyz-order}) and (\ref{tcorr}) of Section 2 by developing an ordering for the turbulent structures in the SOL. In Appendix B.2, we calculate the characteristic steady state gradients parallel to the wall in the magnetic presheath by projecting the SOL width onto the $y$ and $z$ directions parallel to the wall. The steady state lengths are longer than, or of the same order as, the characteristic lengths parallel to the wall due to turbulent structures, as expected.

\subsection{Gradients parallel to the wall and characteristic timescale in the magnetic presheath as a result of turbulent structures in the SOL}

The size of the turbulent structures is assumed of order $l \sim \rho_{\text{i}} / \delta$ in any direction perpendicular to field lines, with $\delta \ll 1$. We proceed to estimate the parallel length $l_{\parallel}$ and turnover time $t_{\text{turn}}$ associated with such structures. %$\partial / \partial t$.
In the perpendicular direction, over a characteristic turbulent timescale, we have assumed that plasma travels a distance $l$. The distance it travels in the parallel direction is larger than $l$ by the factor by which the typical velocity along the field line, the thermal velocity $v_{\text{t,i}}$, is larger than the cross-field one. We order the cross-field velocity of plasma in the SOL the same as the $\vec{E} \times \vec{B}$ drift we expect turbulence to produce, $\sim \delta v_{\text{t,i}}$. Therefore, we can assume that turbulent structures have a size
$l_{\parallel} = l / \delta$
parallel to the magnetic field.
\begin{figure}[h]
\centering
\includegraphics[scale=0.4]{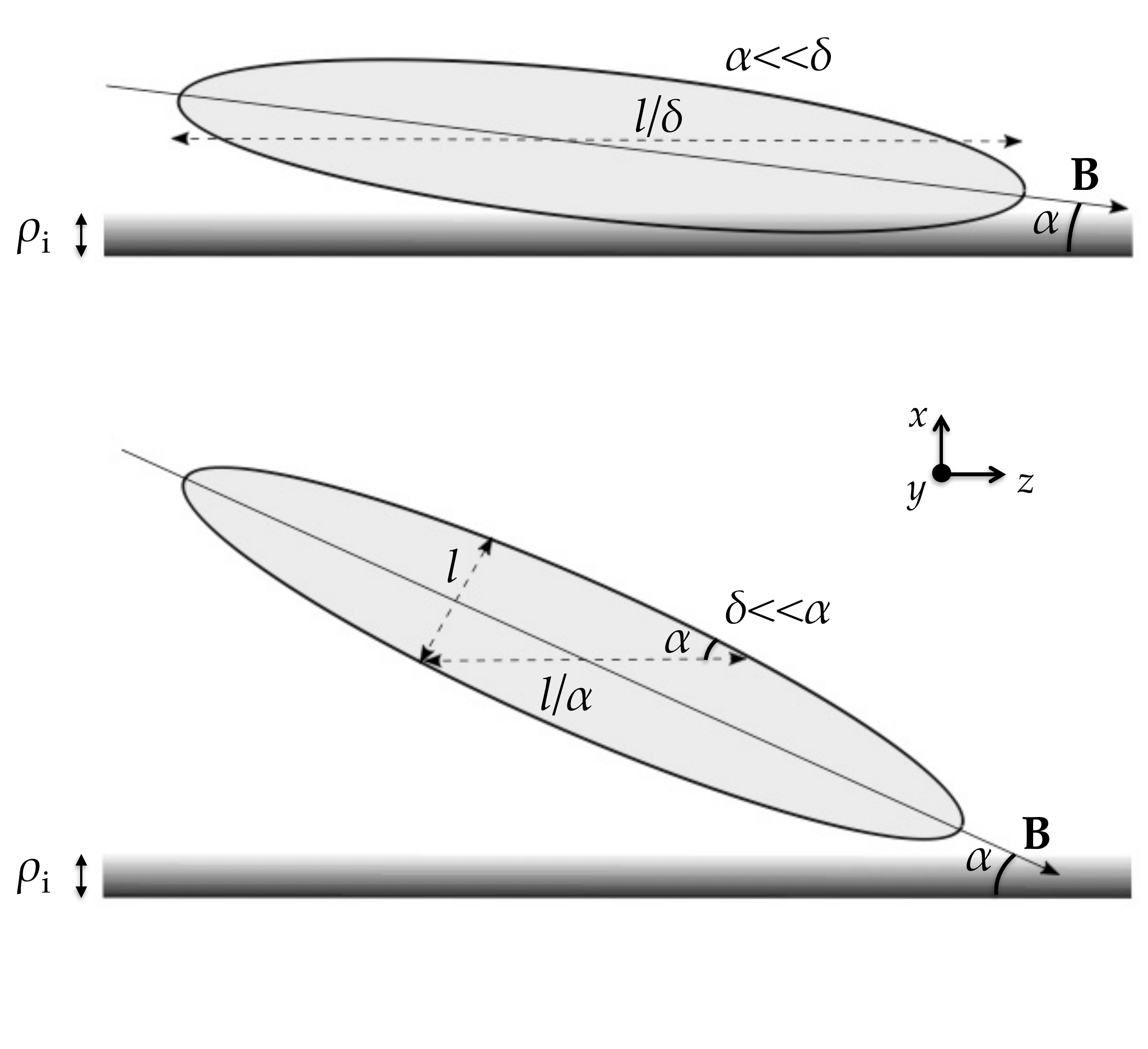}
\caption{Turbulent structures in the SOL, as they approach  the magnetic presheath (thin shaded region of thickness $\sim \rho_{\text{i}}$), are shown here. The elongation of these structures is by a factor $1/\delta$, which comes from the characteristic size of perpendicular velocities compared to parallel ones. Two cases (i) $\alpha \ll \delta$ and (ii) $\delta \ll \alpha$ are shown. In (i), the size of the turbulent structure in the $z$ direction is determined by the length of the turbulent structure parallel to the field line, $l/\delta$. In (ii), it is determined by the length of a cut across the eddy, $l/\alpha$.}
\label{figure-turbulence-size}
\end{figure}

  We refer to turbulent scale lenghs in the $z$ direction as $l_z$. For $\alpha \ll \delta$, it should be clear from Figure \ref{figure-turbulence-size} that gradients in the $z$-direction in the magnetic presheath, arising due to the turbulent structures impinging on the wall, are set by $l/ \delta$, so that $l / l_z \sim \delta$. On the other hand, when $\delta \ll \alpha$ the length scale in the $z$-direction is set by the horizontal cut across the eddy shown in the lower picture, of length $l/ \alpha$, so that $l / l_z \sim \alpha$. Therefore, $l_z \sim \min \left( l / \alpha , l / \delta \right) $. By ordering $z \sim l_z$, we obtain the ordering of (\ref{xyz-order}).
  
  The turnover time of turbulence is obtained from the characteristic length and velocity scales associated with the turbulence,
$t_{\text{turn}} \sim l / \delta v_{\text{t,i}} \sim 1 / \delta ^2 \Omega \text{.}$
 This leads to an estimate for the characteristic frequency of changes within turbulent structures, using $\partial / \partial t \sim 1 / t_{\text{turn}}$, from which the ordering of (\ref{tcorr}) follows.

The gradients in the $x$ direction outside of the magnetic presheath are determined by the cross-field size $l$ of turbulent structures, but they get larger as the magnetic presheath is approached (its characteristic thickness is $\rho_{\text{i}}$). Pictorially, this can be viewed as a squeezing that the turbulent structures undergo in the direction normal to the wall as they approach it. However, the discussion on the characteristic lengths \emph{parallel to the wall} in the magnetic presheath is unaffected, because these scales are inherited from the boundary conditions at the magnetic presheath entrance ($x \rightarrow \infty$).

\subsection{Steady state gradients parallel to the wall}

\begin{figure}[h]
\centering
\includegraphics[scale=0.4]{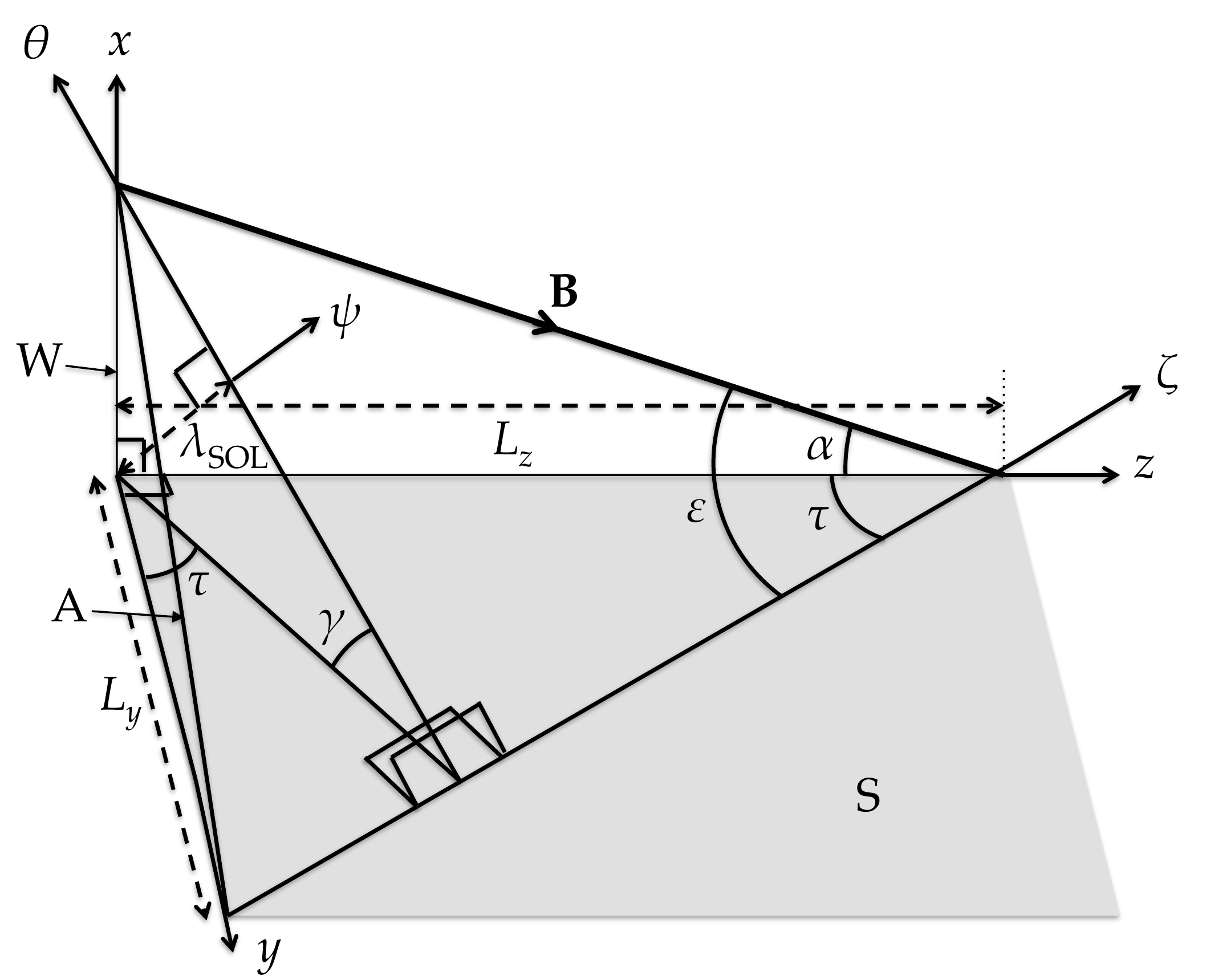}
\caption{A schematic that shows how the SOL thickness $\lambda_{\text{SOL}}$, measured in the flux coordinate direction $\psi$, is projected to lengths $L_y$ and $L_z$ in the directions parallel to the wall. The planes A, W and S and all of the angles, except $\tau$, were introduced in Figure \ref{figure-combined-divertor}. Recall that A is a flux surface that contains magnetic field lines, one of which is shown with a bold arrow and marked $\vec{B}$. The angle $\tau$ is measured between the $z$ and the $\zeta$ (toroidal) directions. }
\label{diagram-SOL-thick}
\end{figure}

The $y$ and $z$ directions are expected to be associated with smaller steady state gradients, as well as turbulent ones, because a component of these directions is in the flux coordinate direction $\psi$ (see Figure \ref{diagram-SOL-thick}), which is associated with the SOL width $\lambda_{\text{SOL}}$.\footnote{The $x$ direction also has a component in the $\psi$ direction, but the scale of the SOL width $\lambda_{\text{SOL}}$ is large compared to the magnetic presheath scale $\rho_{\text{i}}$, so this does not matter.}
Note that, in a typical tokamak, the SOL width is of the order of the width of turbulent structures, $\lambda_{\text{SOL}} \sim l \sim 10 \text{ mm}$ \cite{Carralero-2015}.
We can calculate the projection in the $y$ and $z$ directions of the SOL width, and thus obtain an estimate for the characteristic steady state scale lengths in those directions, $L_{y}$ and $L_{z}$. These must be greater than or equal to the turbulent scales in those directions, $L_y \gtrsim l$ and $L_z \gtrsim l_z \sim \min \left( l/\alpha , l / \delta \right)$.

In Figure \ref{diagram-SOL-thick}, the angle $\varepsilon$ is related to the ratio of the poloidal component of the magnetic field to the toroidal component of the magnetic field, $\tan \varepsilon = \left| B_{\theta} / B_{\zeta} \right| $. The ratio $B_{\theta} / B_{\zeta}$ is usually small in tokamaks, so $\varepsilon \ll 1$. The angle $\gamma$ is the angle between the flux surface and the divertor target. From Figure \ref{diagram-SOL-thick}, we obtain an expression relating $\alpha$ to $\varepsilon$ and $\gamma$, 
\begin{align}
\sin \alpha = \sin\varepsilon \sin\gamma \text{.}\end{align}
In order to achieve $\alpha \ll 1$ it is sufficient to have $\varepsilon \ll 1 $, which is valid for most tokamaks. However, the divertor target inclination in the poloidal plane is a free design parameter, and the flux surface geometry can be controlled with the external magnets. Therefore, the angle $\gamma$ between the divertor target and the flux surface is often also made small, $\gamma \lesssim 1$, in order to make $\alpha$ even smaller. We order $\alpha$ with respect to $\varepsilon$ as $\alpha \lesssim \varepsilon$.

In what follows, it will be convenient to use the angle $\tau$, shown in Figure \ref{diagram-SOL-thick}. We express $\tau$ in terms of $\varepsilon$ and $\gamma$, 
\begin{align} \label{tau-epsilon-gamma}
\tan\tau = \tan \varepsilon \cos\gamma \text{.}
\end{align}
We proceed to express the length scales $L_y$ and $L_z$ in terms of the SOL width $\lambda_{\text{SOL}}$ and the angles $\varepsilon$ and $\gamma$. Projecting the SOL width $\lambda_{\text{SOL}}$ onto the $z$-axis and using (\ref{tau-epsilon-gamma}), we obtain % to express the answer in terms of $\varepsilon$ and $\gamma$)
\begin{align} \label{Lz}
L_z = \frac{\lambda_{\text{SOL}}}{ \sin \gamma \sin \tau }  = \frac{ \sqrt{1 + \tan^2\varepsilon \cos^2 \gamma }}{ \sin \gamma \cos \gamma \tan \varepsilon }  \lambda_{\text{SOL}} \sim \frac{1}{\varepsilon \sin\gamma} \lambda_{\text{SOL}} \gg \lambda_{\text{SOL}}  \text{.}
\end{align}
The presence of $\sin \gamma$ in the denominator of (\ref{Lz}) implies that $L_z \sim \lambda_{\text{SOL}} / \alpha \gtrsim l_z \sim \min \left( l/ \alpha , l / \delta \right) $. The SOL width projected in the $y$ direction is, using (\ref{tau-epsilon-gamma}), 
\begin{align}
L_y = \frac{\lambda_{\text{SOL}}}{\cos \tau \sin \gamma} = \frac{\sqrt{1 + \tan^2 \varepsilon \cos^2 \gamma}}{\sin\gamma} \lambda_{\text{SOL}} \sim \frac{\lambda_{\text{SOL}}}{\sin\gamma} \gtrsim \lambda_{\text{SOL}} \text{.}
\end{align}
This also implies $L_y \gtrsim l$. The ratio of the two steady state length scales is
\begin{align}
\frac{L_y}{L_z} = \tan \varepsilon \cos \gamma \sim \varepsilon \text{.}
\end{align}
This means that $L_z$ is much larger than $L_y$, by a factor $\sim 1 / \varepsilon$.

The length scale $L_y$ is the scale in the $y$ direction over which the gyrokinetic equation (\ref{gyrokinetic-full-explicit}) needs to be solved. The boundary condition (\ref{bc-F0-y}) must be imposed at $y \rightarrow \pm \infty$, which in practice would correspond to a few $L_y$ in both directions.

\section{Large parallel current} \label{appendix:largejz}

In this appendix, we consider the validity of our equations when a large current $\vec{j}^L$ is driven parallel to the magnetic field through the plasma in the magnetic presheath. In Section 2, we ordered the plasma currents using the particle drifts, and the relationship between current components obtained using Maxwell's equations. This means that the ordering we obtained there, $j_z^D \sim \alpha n_{\text{i}} e v_{\text{t,i}} \sim \delta n_{\text{i}} e v_{\text{t,i}} $ is consistent with the piece of the parallel current that flows through the plasma in response to the currents due to the perpendicular drifts. This parallel current is present to satisfy $\nabla \cdot \vec{j}^D = 0$ and maintain charge neutrality, that may otherwise be broken by the divergence of the perpendicular current. We find that our equations are consistent with a larger parallel current, $j^L \gg \delta e n_{\text{i}} v_{\text{t,i}}$, provided that the size of this current does not become too large, at which point the neglect of the plasma produced magnetic fields and the induced electric fields would no longer be valid assumptions. 

The large parallel current $\vec{j}^L$ has components $j^L_z = j^L \cos \alpha \sim j^L$, $j_x^L = - j^L \sin \alpha  \sim \alpha j^L $ and $j_y^L = 0$. This parallel current must satisfy $\nabla \cdot \vec{j}^L = 0$. Using (\ref{xyz-order}) we find $\partial j^L_x / \partial x \sim \alpha j^L / \rho_{\text{i}}$ and $\partial j^L_z / \partial z \sim \delta \alpha j^L / \rho_{\text{i}} \sim \delta^2 j^L / \rho_{\text{i}}$. Thus, $\nabla \cdot \vec{j}_L = 0$ requires $\partial j^L / \partial x = 0$, and to lowest order $\vec{j}^L$ is not affected by the magnetic presheath.
Therefore, the length scale in the $x$ direction of the large parallel current and the magnetic and electric fields associated with it is larger than the magnetic presheath scale $\rho_{\text{i}}$. Balancing $\partial j_x^L / \partial x \sim \partial j_z^L / \partial z$ leads to ordering 
\begin{align}
\frac{\partial}{\partial x} \sim \frac{\delta}{\alpha l} \sim \frac{1}{l} \text{,}
\end{align}
so the appropriate length scale in the $x$ direction (for this appendix) is the perpendicular turbulent scale $l$.

The magnetic field produced in the plasma by $\vec{j}^L$ is denoted $\vec{B}^{p'}$ and is determined by Amp\`ere's law (\ref{Ampere}), with $\vec{j}^L$ instead of $\vec{j}^D$ and $\vec{B}^{p'}$ instead of $\vec{B}^{p}$. Taking the $y$ component of (\ref{Ampere}) with $j_y^L = 0$, we have $\partial B_z^{p'} / \partial x \sim \partial B_x^{p'} / \partial z$, from which we obtain $B_z^{p'} \sim \alpha B_x^{p'}$. Considering the long length scales in the $z$ direction, this implies that $\partial B_z^{p'} / \partial z$ must be subdominant in $\nabla \cdot \vec{B}^{p'} = 0$, so $\partial B_x^{p'} / \partial x \sim \partial B_y^{p'} / \partial y $. This implies that $B_x^{p'} \sim B_y^{p'}$. The $x$ and $z$ components of Amp\`ere's law determine $B_y^{p'} \sim B_x^{p'} \sim \mu_0 l j^L $. Collecting the orderings for the components of the magnetic field produced by the large parallel current, we have
\begin{align} \label{ordering-Bx-By-Bz-largejz}
B_z^{p'} \sim \alpha B^{p'} \ll B_x^{p'} \sim B_y^{p'} \sim B^{p'} \sim \mu_0 l j^L \sim \frac{ j^L }{\delta en_{\text{i}}v_{\text{t,i}}} \beta  B^c \text{,}
\end{align}
where in the rightmost equation we used $\beta B^c \sim B^p \sim \mu_0 l \delta n_{\text{i}} e v_{\text{t,i}} $ inferred from (\ref{j-orderings}) and (\ref{ordering-beta}).

As explained in Section 2, in order to neglect $\vec{B}^{p'}$ we require each component of it to be negligible compared to either the respective component or the smallest retained component of the constant external magnetic field $\vec{B}^c$. The strongest constraint is obtained by the neglect of $B_x^{p'}$ and $B_y^{p'}$ compared to $B_x^c \sim \alpha B^c$. This is $B^{p'} \ll \alpha B^c$, which leads to
\begin{align} \label{ordering-large-current}
j^L \ll \left( \frac{\alpha}{\beta} \right) \delta e n_{\text{i}} v_{\text{t,i}} \text{.}
\end{align}

The large parallel current $\vec{j}^L$ is consistent with an electrostatic electric field provided that each component of the electric field $\vec{E}^{p'}$ induced by $\vec{B}^{p'}$ is negligible compared to either the respective or the smallest retained component of the electrostatic electric field, $-\nabla \phi$. From (\ref{tcorr}), (\ref{ordering-Bx-By-Bz-largejz}) and the length scale orderings of this section (which are $x \sim y \sim l \ll z \sim l / \delta \sim l / \alpha$), we can order the components of the induction equation (\ref{induction}), with $\vec{E}^{p'}$ instead of $\vec{E}^{p}$ and $\vec{B}^{p'}$ instead of $\vec{B}^p$. We obtain (recalling that $\rho_{\text{i}} \Omega = v_{\text{t,i}}$)
\begin{align}
E_x^{p'} \sim E_y^{p'} \sim \alpha E^{p'} \ll E_z^{p'} \sim E^{p'} \sim \delta v_{\text{t,i}} B^{p'} \text{.}
\end{align}
Therefore we find that the strongest constraint on the electrostatic approximation is $E^{p'}_z \sim \delta v_{\text{t,i}} B^{p'} \ll \delta v_{\text{t,i}} B^c$ and leads to $B^{p'} \ll B^c$. This is a weaker condition than the one needed to neglect the magnetic field $\vec{B}^{p'}$, hence the electrostatic approximation is justified when (\ref{ordering-large-current}) holds. Note that (\ref{ordering-large-current}) allows for a large parallel current in the magnetic presheath because $\alpha \gg \beta$, as pointed out in Section 2. If currents larger than (\ref{ordering-large-current}) were present in the magnetic presheath, we would have to consider the magnetic fields produced by them in our equations (and in extreme cases also the induced electric fields). For example, currents larger than (\ref{ordering-large-current}) would be large enough to change the angle $\alpha$ between the magnetic field lines and the wall.

\section{Calculating the adiabatic invariant} \label{appendix:mucalc}

In this section we prove that the lowest order adiabatic invariant has the form in (\ref{mu-gyroav}), starting from the definition
\begin{align} \label{adiabatic-definition}
\mu = \frac{1}{2\pi m_{\text{i}}}\oint  \vec{\tilde{p}} \cdot \vec{d\tilde{r}} \text{,}
\end{align}
where $\vec{\tilde{p}}$ and $\vec{\tilde{r}}$ are the canonical momentum  and the position vector of the charged particle in the frame where the motion is periodic. We work using the particle equations of Section 3, valid for $\alpha = 0$ and $\delta = 0$. The integral $\oint$ in (\ref{adiabatic-definition}) is performed over one orbit. The transformation to the frame where the ion motion is periodic is obtained by subtracting the gyrophase independent piece of $\vec{v}$ from itself, $\vec{\tilde{v}} = \vec{v} - \left\langle \vec{v} \right\rangle_{\varphi}$. The position $\vec{\tilde{r}}$ is obtained by integrating $\vec{\tilde{v}}$ in time (which, due to periodicity, is equivalent to integrating in gyrophase and dividing by $\overline{\Omega}$). For simplicity, we choose $\vec{\tilde{r}}$ such that $\left\langle \vec{\tilde{r}} \right\rangle_{\varphi} = 0$. From the zeroth order equations (\ref{vx-f-gyro}), (\ref{vy-reexpressed}) and (\ref{vz-U-Uperp}), we obtain
\begin{align} \label{v-gyroaveraged}
\left\langle \vec{v} \right\rangle_{\varphi} =
\begin{pmatrix}
0 \\
\frac{1}{B} \left\langle \frac{\partial \phi}{\partial x} \left(x_{\varphi} \right) \right\rangle_{\varphi} \\
v_{\parallel} \left( U_{\perp}, U, \sigma_{\parallel} \right) \end{pmatrix} \text{.}
\end{align}
An example of translating to the frame where ion motion is periodic is given, for the linear electric field, in Figure \ref{figure-linear-E-effect}. Equation (\ref{v-gyroaveraged}) indicates clearly that the ion orbit is comprised, in addition to the periodic motion, of an $\vec{E} \times \vec{B}$ drift in the $y$ direction and parallel streaming in the $z$ direction.

The electric field in the frame where motion is periodic is given by $\vec{\tilde{E}} = \vec{E} - \left\langle \vec{E} \right\rangle_{\varphi}$, whereas the magnetic field is unchanged provided we are in the non-relativistic limit. From (\ref{mag-vec-pot}), the magnetic vector potential can be expressed as (using $\alpha = 0$).
\begin{align}
\vec{A} = \begin{pmatrix}
0 \\
\tilde{x}B \\ 0
\end{pmatrix} \text{.}
\end{align}
The canonical momentum is $\vec{\tilde{p}} = m_{\text{i}} \vec{\tilde{v}} + Ze\vec{A}$, which leads to the expression for the adiabatic invariant in (\ref{adiabatic-definition}) becoming (using $\vec{d\tilde{r}} = \vec{\tilde{v}} d\varphi / \overline{\Omega} $)
\begin{align}
\mu = \frac{1}{2\pi \overline{\Omega}} \oint \left( \tilde{v}^2 + \Omega \tilde{x} \tilde{v}_y \right) d \varphi \text{.}
\end{align}
From (\ref{vy-xbar-def}) we extract $\tilde{x} = - \tilde{v}_y / \Omega$ which leads to (using $\tilde{v}_x = v_x$)
\begin{align} \label{inv2}
\mu = \frac{1}{2\pi \overline{\Omega}} \oint v_x^2 ~d\varphi \text{.}
\end{align}
Using the definition of a gyroaverage in (\ref{gyroav-def}), equation (\ref{inv2}) reduces to the form in (\ref{mu-gyroav}).

\section{The zeroth order problem with a linear electric field} \label{appendix:linE}

To obtain some physical insight into lowest order orbit distortion, we solve the zeroth order problem ($\alpha = 0$, $\delta = 0$) in a linearly varying electric field and then extract the most obvious physical effects from the calculation. This problem is also solved in reference \cite{Claassen-Gerhauser-1996}. 

Let the electric field of equation (\ref{vx-EOM}) be given by
\begin{align} \label{lin-E}
- \frac{d \phi}{d x} = - \bar{E} + E'x 
\end{align}
with $\bar{E}$ and $E'$ constants. Note that if we take $\bar{E}$ and $E'$ both positive and only look at the region where both $x$ and $d \phi /d x $ are positive, the electric field is directed towards $x=0$ (the wall) and increasing as $x$ gets smaller, which is qualitatively similar to the electric field in the magnetic presheath. In this section, we use equations (\ref{Omegabar-def}), (\ref{varphi-def}), (\ref{xgk-def}),  (\ref{vx-f-gyro}) and (\ref{vy-reexpressed}), which are valid for a general electric field, in order to solve the zeroth order problem with the linear electric field (\ref{lin-E}).

The effective potential in the linear electric field is
\begin{align} \label{chi-linear}
\chi\left(x, \bar{x}\right) = \frac{1}{2}\Omega^2\left(x-\bar{x}\right)^2 + \frac{Ze\bar{E}}{m_{\text{i}}}x - \frac{ZeE'}{2m_{\text{i}}}x^2 \text{.}
\end{align}
Collecting terms multiplying different powers of $x$, we get 
\begin{align} \label{chi-linear-2}
\chi \left( x, \bar{x} \right) = \left( \frac{1}{2} \Omega^2 - \frac{ZeE'}{2m_{\text{i}}} \right) x^2 - \left( \Omega^2 \bar{x} - \frac{Ze\bar{E}}{m_{\text{i}}} \right) x + \frac{1}{2} \Omega^2 \bar{x}^2 \text{,}
\end{align}
which leads to 
\begin{align} \label{chi-linear-3}
2\left( \chi \left( x, \bar{x} \right) - U_{\perp} \right) = \left(  \Omega^2 - \frac{ZeE'}{m_{\text{i}}} \right) x^2 - 2\left( \Omega^2 \bar{x} - \frac{Ze\bar{E}}{m_{\text{i}}} \right) x - 2U_{\perp} + \Omega^2 \bar{x}^2 \text{.}
\end{align}
We factorize $\Omega^2$ and complete the square in (\ref{chi-linear-3}) to get
\begin{align} \label{chi-linear-4}
2\left( \chi \left( x, \bar{x} \right) - U_{\perp} \right) = \Omega^2 \left(  1 - \frac{E'}{\Omega B} \right) \left[ \left( x - \frac{\bar{x} - \bar{E} / \Omega B }{ 1 - E' / \Omega B} \right)^2 - A^2  \right] \text{,}
\end{align}
where we introduced the quantity $A$, which has dimensions of length and is given by
\begin{align} \label{A-Uperp-xbar}
A^2 = \left( \frac{\bar{x} - \bar{E} / \Omega B }{ 1 - E' / \Omega B} \right)^2 + \frac{ 
2U_{\perp} / \Omega^2 -  \bar{x}^2}{ 1 - E' / \Omega B } \text{.}
\end{align}

We manipulate the integral in (\ref{Omegabar-def}) to 
\begin{align} \label{integral-1}
\frac{2\pi}{\overline{\Omega}} = \frac{2}{\Omega\sqrt{1 - \frac{E'}{\Omega B}}} \int_{x_{\text{b}}}^{x_{\text{t}}} \frac{dx}{\sqrt{ A^2 - \left(x - \frac{\bar{x} - \bar{E} / \Omega B }{ 1 - E' / \Omega B}\right)^2}} \text{,}
\end{align}
with $x_{\text{b}}$ and $x_{\text{t}}$ given by
\begin{align}
x_{\text{b}} = \frac{\bar{x} - \bar{E} / \Omega B }{ 1 - E' / \Omega B} - A \text{,}
\end{align}
\begin{align}
x_{\text{t}} = \frac{\bar{x} - \bar{E} / \Omega B }{ 1 - E' / \Omega B} + A \text{.}
\end{align}
We carry out the integral in (\ref{integral-1}) to obtain
\begin{align} \label{Omegabar-linear-E}
\overline{\Omega} = \Omega \sqrt{1 - \frac{E'}{\Omega B}} \text{.}
\end{align}
The gyrofrequency is altered by the gradient of the electric field $E'$, and the condition for periodicity is set by $\overline{\Omega}$ being real, that is
\begin{align}
\frac{E'}{\Omega B} < 1 \text{.}
\end{align}
We note that the periodic motion is lost if $E' / \Omega B \geq 1$, which can only happen for positive $E'/Z$, that is for a diverging (converging) electric field applied on a positive (negative) charge. The case without periodic motion corresponds to a quadratic effective potential that has a maximum but no minimum. 

From (\ref{varphi-def}) we have, upon using the expressions (\ref{chi-linear-4}) and (\ref{Omegabar-linear-E}),
\begin{align}
\varphi = \sigma_x \int_{x_{\text{t}}}^{x} \frac{dx'}{\sqrt{ A^2 - \left(x' - \left( \Omega/\overline{\Omega} \right)^{2} \left(  \bar{x} - \bar{E}/\Omega B\right) \right)^2}}
\text{.}
\end{align}
This can be solved to get
\begin{align} \label{x-linear-unsimp}
x = x_{\text{gk}} \left( \bar{x}, U_{\perp}, \varphi \right) \equiv A\cos\varphi  + \left( \frac{\Omega}{\overline{\Omega}} \right)^2 \left( \bar{x} - \frac{\bar{E}}{\Omega B} \right) \text{,}
\end{align}
which can also be written as
\begin{align} \label{x-linear}
x_{\text{gk}} \left( \bar{x}, U_{\perp} , \varphi \right) = A\cos\varphi + \bar{x}  - \left( \frac{\Omega}{\overline{\Omega}} \right)^2 \frac{1}{\Omega B}\frac{d \phi}{dx}\left(\bar{x}\right) \text{.}
\end{align}
Equations (\ref{x-linear-unsimp}) and (\ref{x-linear}) correspond to $x_{\text{gk}} \left( \bar{x}, U_{\perp}, \varphi \right)$ in  (\ref{xgk-def}) for the linear electric field. The quantity $A$ given in (\ref{A-Uperp-xbar}) is the amplitude of the particle orbit in the $x$ direction. Note that gyroaveraging (\ref{x-linear}) gives
\begin{align} \label{xgyro-xbar-linear}
\left\langle x \right\rangle_{\varphi} = \bar{x} - \left( \frac{\Omega}{\overline{\Omega}} \right)^2 \frac{1}{\Omega B} \frac{d \phi}{dx} \left( \bar{x} \right) \text{,}
\end{align}
which gives the difference between the orbit position, $\bar{x}$, and the gyroaverage of the particle $x$ coordinate over the orbit (the conventional guiding centre), $\left\langle x \right\rangle_{\varphi}$, for the linear electric field configuration. 

From (\ref{vx-f-gyro}) we have
\begin{align} \label{vx-linear}
v_{x} = - \overline{\Omega} A\sin \varphi \text{,}
\end{align}
and using (\ref{vy-reexpressed}) (or directly from (\ref{vy-xbar-def}) and (\ref{x-linear})) we obtain
\begin{align} \label{vy-linear}
v_{y} = - \Omega A \cos \varphi  + \left( \frac{\Omega}{\overline{\Omega}} \right)^2 \frac{1}{B}\frac{d\phi}{dx}\left(\bar{x}\right) \text{.}
\end{align}
Note that 
\begin{align} \label{linear-el-field-relation}
\left\langle \frac{d \phi}{d x} \left( x_{\text{gk}} \left( \bar{x}, U_{\perp}, \varphi \right) \right) \right\rangle_{\varphi} =  \frac{d \phi}{dx} \left( \left\langle x \right\rangle_{\varphi} \right) = \left( \frac{\Omega}{\overline{\Omega}} \right)^2 \frac{d \phi}{dx}\left(\bar{x}\right) \text{,}
\end{align}
where we have used (\ref{xgyro-xbar-linear}). Thus, the second term on the RHS of (\ref{vy-linear}) is just the gyroaverage of the electric field divided by $B$, which is the $\vec{E}\times \vec{B}$ drift $\left\langle \dot{y} \right\rangle_{\varphi}$. Equations (\ref{vx-linear}) and (\ref{vy-linear}) are the equivalent of (\ref{vx-f-gyro}) and (\ref{vy-reexpressed}) for the linear electric field.
The velocity in the $z$ direction is given by the expression (\ref{vz-U-Uperp}). 

\begin{figure}[h!]
\centering
\includegraphics[scale=0.4]{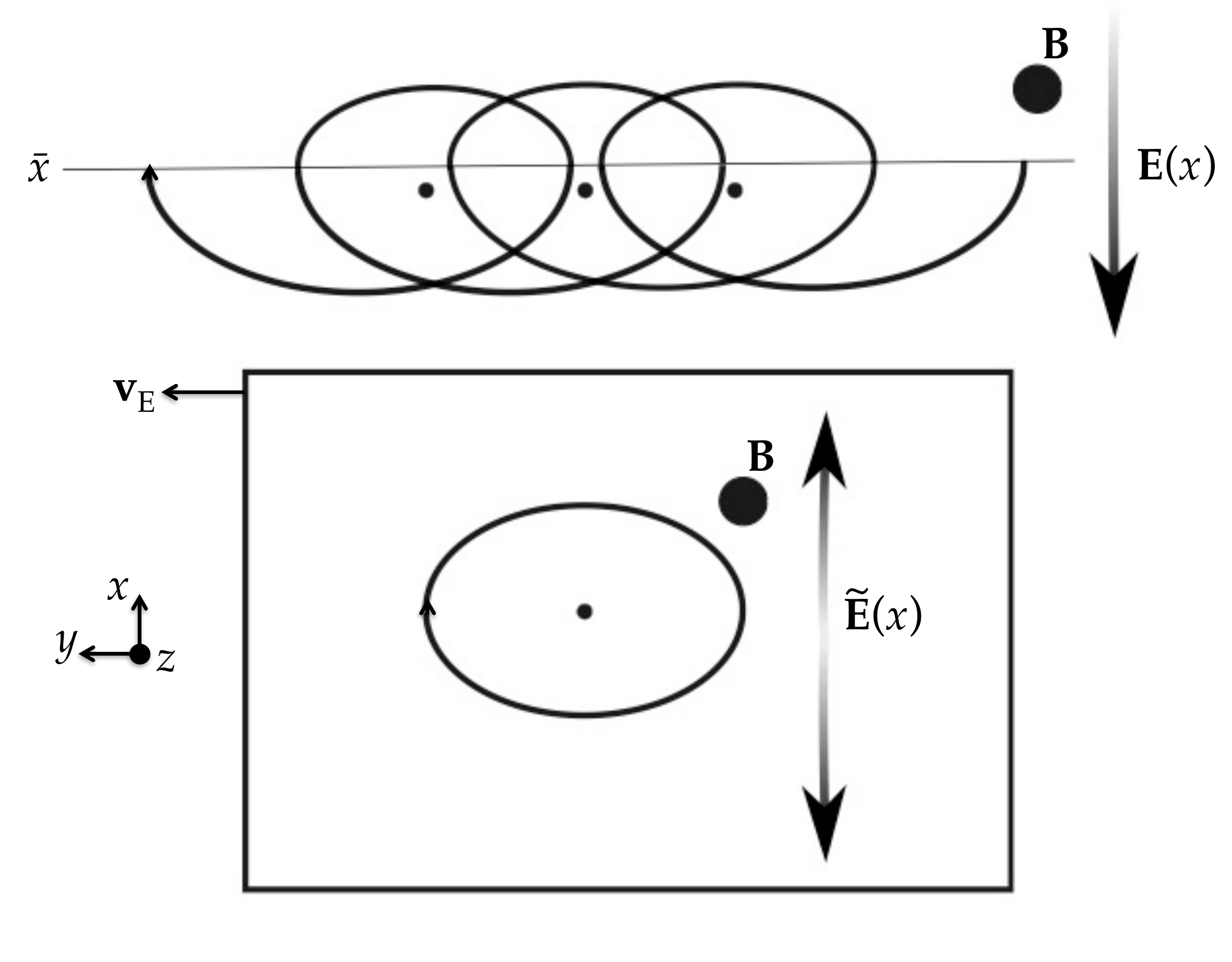}
\caption{Orbit squeezing in a linear electric field, shown in two different frames of reference: with (above) and without (below) $\vec{E}\times\vec{B}$ drift, labelled $\vec{v}_{\text{E}}$. The magnetic field $\vec{B}$ is marked by a large dot and is directed out of the page. The electric field $\vec{E}\left(x\right)$ is marked by arrows and is denoted with a tilde in the drifting frame (below). The gyroaverage of the $x$ coordinate, $\left\langle x \right\rangle_{\varphi}$, is marked by a small dot in both frames of reference. There is a difference between $\bar{x}$, shown by the horizontal line in the upper diagram, and $\left\langle x \right\rangle_{\varphi}$, given by (\ref{xgyro-xbar-linear}).}
\label{figure-linear-E-effect}
\end{figure}

The helical motion is modified from being circular in the $x$-$y$ plane to being an ellipse in the frame of reference where the $\vec{E} \times \vec{B}$ drift is zero, as shown in Figure \ref{figure-linear-E-effect}. This is because, from integrating (\ref{vy-linear}) in time (using $\varphi = \overline{\Omega} t $), the amplitude of the ion orbit in the $y$ direction is $\Omega A / \overline{\Omega}$, which in general is not equal to $A$. For the case $E'/Z > 0$, the amplitude in the $y$ direction is larger. The electric field squeezes the orbit in the $x$ direction compared to the $y$ direction because it increases the local radius of curvature of the trajectory of the particle by increasing its orbital speed (see Figure \ref{figure-linear-E-effect}, bottom picture).

The adiabatic invariant for an ion moving in the linear electric field configuration is, from (\ref{mu-gyroav}) and (\ref{vx-linear}),
\begin{align} \label{mu-linear}
\mu = \frac{1}{2}\overline{\Omega} A^2 \text{.}
\end{align}
This means that we can express $A$ as a function of $\mu$ only,
\begin{align} \label{A-mu}
A = \sqrt{\frac{2\mu}{\overline{\Omega}}} \text{.}
\end{align}
The form of the adiabatic invariant (\ref{mu-linear}) implies that if one were to slowly decrease $\overline{\Omega}$ to zero by increasing the electric field gradient $E'$, the amplitude of the ellipse in the $x$ direction (semi-minor axis), $A$, would increase as $1/\overline{\Omega}^{1/2}$. However, the amplitude in the $y$ direction (semi-major axis), $\Omega A / \overline{\Omega}$, would increase even more, as $1/\overline{\Omega}^{3/2}$, so the orbit would increase in size and, at the same time, become more squeezed. Eventually the orbit would become infinitely squeezed and large, and open up when $\overline{\Omega}^2 \leq 0$. This corresponds to a flattening of the (parabolic) effective potential curve until eventually the minimum turns into a maximum.

\section{Derivatives of $\mu$ used in Section 4.3} \label{appendix:muderivatives}

In this appendix, we derive equations (\ref{dmudxbar}), (\ref{dmudY}) and (\ref{dmudUperp}). The expression for the magnetic moment is, including the $y$ dependence,
\begin{align} \label{mu(Uperp,xbar,y)-appendix}
\mu = \mu_{\text{gk}} \left(\bar{x}, y , U_{\perp} \right) = \frac{1}{\pi} \int_{x_{\text{b}}}^{x_{\text{t}}} \sqrt{2 \left( U_{\perp} - \chi \left( x, \bar{x}, y \right) \right)} dx \text{.}
\end{align}
Differentiating with respect to $\bar{x}$ gives, using $\partial \chi / \partial \bar{x} = \Omega^2 \left( \bar{x} - x \right) = \Omega v_y$, equation (\ref{vy-reexpressed}) and the definition of the gyroaverage (\ref{gyroav-def}),
\begin{align}
\frac{\partial \mu}{\partial \bar{x}} = \frac{1}{\pi} \int_{x_{\text{b}}}^{x_{\text{t}}} \frac{\Omega^2 \left( x - \bar{x} \right)}{\sqrt{2\left( U_{\perp} - \chi\left( x, \bar{x}, y \right) \right)}}dx = -\frac{Ze}{m_{\text{i}} \overline{\Omega}} \left\langle \frac{\partial \phi}{\partial x} \left( x_{\varphi}, y \right) \right\rangle_{\varphi} \text{.}
\end{align}
This is equation (\ref{dmudxbar}).
Differentiating (\ref{mu(Uperp,xbar,y)-appendix}) with respect to $y$ we obtain, using $\partial \chi / \partial y = \left( Ze /m_{\text{i}} \right) \partial \phi / \partial y$
\begin{align}
\frac{\partial \mu}{\partial y} = - \frac{1}{\pi} \frac{Ze}{m_{\text{i}}} \int_{x_{\text{b}}}^{x_{\text{t}}} \frac{\partial \phi / \partial y \left(x, y \right)}{\sqrt{2\left( U_{\perp} - \chi\left( x, \bar{x}, y \right) \right)}} dx = -\frac{Ze}{m_{\text{i}} \overline{\Omega}} \left\langle \frac{\partial \phi}{\partial y} \left( x_{\varphi}, y \right) \right\rangle_{\varphi} \text{.}
\end{align}
This recovers equation (\ref{dmudY}). Finally, differentiating (\ref{mu(Uperp,xbar,y)-appendix}) with respect to $U_{\perp}$ and using (\ref{Omegabar-def}) we have equation (\ref{dmudUperp}),
\begin{align}
\frac{\partial \mu}{\partial U_{\perp}} = \frac{1}{\pi} \int_{x_{\text{b}}}^{x_{\text{t}}} \frac{1}{\sqrt{2\left( U_{\perp} - \chi\left( x, \bar{x}, y \right) \right)}} dx = \frac{1}{\overline{\Omega}} \text{.}
\end{align}

\section*{References}

\bibliography{gyrokineticsbibliography}{}
\bibliographystyle{unsrt}

\end{document}